%% file: arxiv_main.tex
\documentclass{article}

\usepackage{arxiv}

\input{packages}

\newcommand{\sys}{\textsc{SACTOR}\xspace}

\title{\sys{}: LLM-Driven Correct and Idiomatic C to Rust Translation with Static Analysis and FFI-Based Verification}

\input{author_stuff}

\renewcommand{\headeright}{}
\renewcommand{\undertitle}{}

\date{}

\begin{document}

\maketitle

\input{contents/0_abstract}


\input{contents/1_introduction}
\input{contents/2_background}
\input{contents/3_related_works}
\input{contents/5_method_details}
\input{contents/6_experimental_setup}
\input{contents/7_evaluation}

\input{contents/8_conclusion}

\input{contents/limitations}

\newpage

\bibliographystyle{unsrtnat}
\bibliography{references}

\newpage
\appendix

\input{contents/appendix}

\end{document}

%% file: packages.tex
\usepackage{graphicx}
\usepackage[dvipsnames,x11names]{xcolor}
\usepackage{booktabs} %
\usepackage{tabularx}
\usepackage{longtable}
\usepackage{nameref}
\usepackage{enumitem}
\usepackage{soul} %
\usepackage{algpseudocode} %
\usepackage{pifont}    %
\usepackage{bbding}
\usepackage{array}
\usepackage{colortbl}  %
\usepackage{listings-rust}
\usepackage{algorithm}
\usepackage{algorithmicx}
\usepackage{xspace}
\usepackage{subcaption}
\usepackage[T1]{fontenc}
\usepackage[utf8]{inputenc}
\usepackage{times}
\usepackage{latexsym}
\usepackage{microtype}
\usepackage{inconsolata}
\usepackage{adjustbox}
\usepackage{afterpage}
\usepackage{float}
\usepackage{threeparttable}
\usepackage{hyperref}
\usepackage{natbib}

\definecolor{codegreen}{rgb}{0,0.6,0}
\definecolor{codegray}{rgb}{0.5,0.5,0.5}
\definecolor{codepurple}{rgb}{0.58,0,0.82}
\definecolor{backcolor}{rgb}{0.95,0.95,0.92}
\definecolor{backcolor1}{HTML}{f6f6f6}
\definecolor{backcolor2}{HTML}{fcf4e2}
\definecolor{backcolor3}{HTML}{eff1f5}

\usepackage{tcolorbox}
\tcbuselibrary{listings}

\newenvironment{myitemize}
  {\begin{itemize}[leftmargin=1em]} %
  {\end{itemize}}

\lstdefinestyle{mystyle}{
    backgroundcolor=\color{backcolor},
    commentstyle=\color{codegreen},
    keywordstyle=\color{magenta},
    stringstyle=\color{codepurple},
    basicstyle=\ttfamily\small,%
    breakatwhitespace=false,
    breaklines=true,
    captionpos=b,
    keepspaces=true,
    numbersep=5pt,
    showspaces=false,
    showstringspaces=false,
    showtabs=false,
    tabsize=2,
    lineskip=-1pt
}
\newcommand{\roundedinclude}[3][]{%
    \begin{tcolorbox}[
        arc=1.5mm,
        colback=#1,
        boxrule=0.5pt,
        left=1mm,
        right=1mm,
        top=0mm,
        bottom=0mm
    ]
    \includegraphics[width=#3]{#2}
    \end{tcolorbox}
}

\let\oldComment\Comment
\renewcommand{\Comment}[1]{\textcolor{gray}{\oldComment{#1}}}

\algrenewcommand\algorithmicindent{1em} %

\usepackage{amsmath}
\usepackage{amssymb}
\usepackage{mathtools}
\usepackage{amsthm}

\usepackage[capitalize,noabbrev]{cleveref}

\theoremstyle{plain}

\theoremstyle{definition}

\theoremstyle{remark}

\usepackage[textsize=small]{todonotes}

\newenvironment{squishitemize}
{\begin{list}{\textbullet}{%
    \setlength{\itemsep}{0pt}%
    \setlength{\parsep}{0pt}%
    \setlength{\topsep}{0pt}%
    \setlength{\parskip}{0pt} %
    \setlength{\labelwidth}{.5in}%
    \setlength{\labelsep}{0.05in} %
    \setlength{\leftmargin}{.15in} %
    }}
  {\end{list}}

  \newenvironment{squishenumerate}
  {\begin{list}{\arabic{enumi}.}{%
    \usecounter{enumi}%
    \setlength{\itemsep}{0pt}%
    \setlength{\parsep}{0pt}%
    \setlength{\topsep}{0pt}%
    \setlength{\parskip}{0pt}%
    \setlength{\labelwidth}{.5in}%
    \setlength{\labelsep}{0.05in}%
    \setlength{\leftmargin}{.2in}}}
  {\end{list}}

%% file: author_stuff.tex
\usepackage{authblk}

\author[1]{%
    \hspace{1mm}Tianyang Zhou\thanks{\texttt{tz64@illinois.edu}}
}
\author[2]{%
    \hspace{1mm}Ziyi Zhang\thanks{\texttt{ziyi.zhang2@wisc.edu}}
}
\author[1]{%
    \hspace{1mm}Haowen Lin\thanks{\texttt{haowenl3@illinois.edu}}
}
\author[2]{%
    \hspace{1mm}Somesh Jha\thanks{\texttt{jha@cs.wisc.edu}}
} 
\author[3]{%
    \hspace{1mm}Mihai Christodorescu\thanks{\texttt{christodorescu@google.com}}
}
\author[1]{
    \hspace{1mm}Kirill Levchenko\thanks{\texttt{klevchen@illinois.edu}}
}
\author[1]{
    \hspace{1mm}Varun Chandrasekaran\thanks{\texttt{varunc@illinois.edu}}
}

\affil[1]{University of Illinois Urbana-Champaign}
\affil[2]{University of Wisconsin--Madison}
\affil[3]{Google}

%% file: contents/0_abstract.tex
\begin{abstract}

Translating software written in C to Rust has significant benefits in improving memory safety. However, manual translation is cumbersome, error-prone, and often produces unidiomatic code. Large language models (LLMs) have demonstrated promise in producing idiomatic translations, but offer no correctness guarantees. We propose \sys{}, an LLM-driven C-to-Rust translation tool that employs a two-step process: an initial ``unidiomatic'' translation to preserve interface, followed by an ``idiomatic'' refinement to align with Rust standards. %
To validate correctness of our function-wise incremental translation that mixes C and Rust, we use end-to-end testing via the foreign function interface.
We evaluate \sys{} on $200$ programs from two public datasets and on two more complex scenarios (a 50-sample subset of CRust-Bench and the \texttt{libogg} library), comparing multiple LLMs.
Across datasets, \sys{} delivers high end-to-end correctness and produces safe, idiomatic Rust with up to 7$\times$ fewer Clippy warnings;
On CRust-Bench, \sys{} achieves an average (across samples) of 85\% unidiomatic and 52\% idiomatic success, and on \texttt{libogg} it attains full unidiomatic and up to 78\% idiomatic coverage on GPT-5.

\end{abstract}

%% file: contents/1_introduction.tex
\section{Introduction}
\label{sec:intro}

C is widely used due to its ability to directly manipulate memory and hardware~\citep{love2013linux}. However, manual memory management leads to vulnerabilities such as buffer overflows, dangling pointers, and memory leaks~\citep{bigvul}. Rust addresses these issues by enforcing memory safety through a strict ownership model without garbage collection~\citep{matsakis2014rust}, and has been adopted in projects like the Linux kernel\footnote{\url{https://github.com/Rust-for-Linux/linux}} and Mozilla Firefox.
{\em Translating legacy C code into idiomatic Rust improves safety and maintainability, but manual translation is error-prone, slow, and requires expertise in both languages.}

Automatic tools such as C2Rust~\citep{c2rust} generate Rust by analyzing C ASTs, but rule-based or static approaches~\citep{crown, c2rust, emre2021translating, hong2024don, ling2022rust} typically yield unidiomatic code with heavy use of \lstinline!unsafe!. Given semantic differences between C and Rust, idiomatic translations are crucial for compiler-enforced safety, readability, and maintainability.

Large language models (LLMs) show potential for capturing syntax and semantics~\citep{pan2023understanding}, but they hallucinate and often generate incorrect or unsafe code~\citep{perry2023users}. In C-to-Rust translation, naive prompting produces unsafe or semantically misaligned outputs. Prior work has explored prompting strategies~\citep{syzygy, c2saferrust, shiraishi2024context} and verification methods such as fuzzing and symbolic execution~\citep{vert, flourine}. While these improve correctness, they struggle with complex programs and rarely yield idiomatic Rust. For example, Vert~\citep{vert} fails on programs with complex data structures, and C2SaferRust~\citep{c2saferrust} still produces Rust with numerous \lstinline!unsafe! blocks.

\begin{figure}
  \centering
  \includegraphics[width=0.7\linewidth]{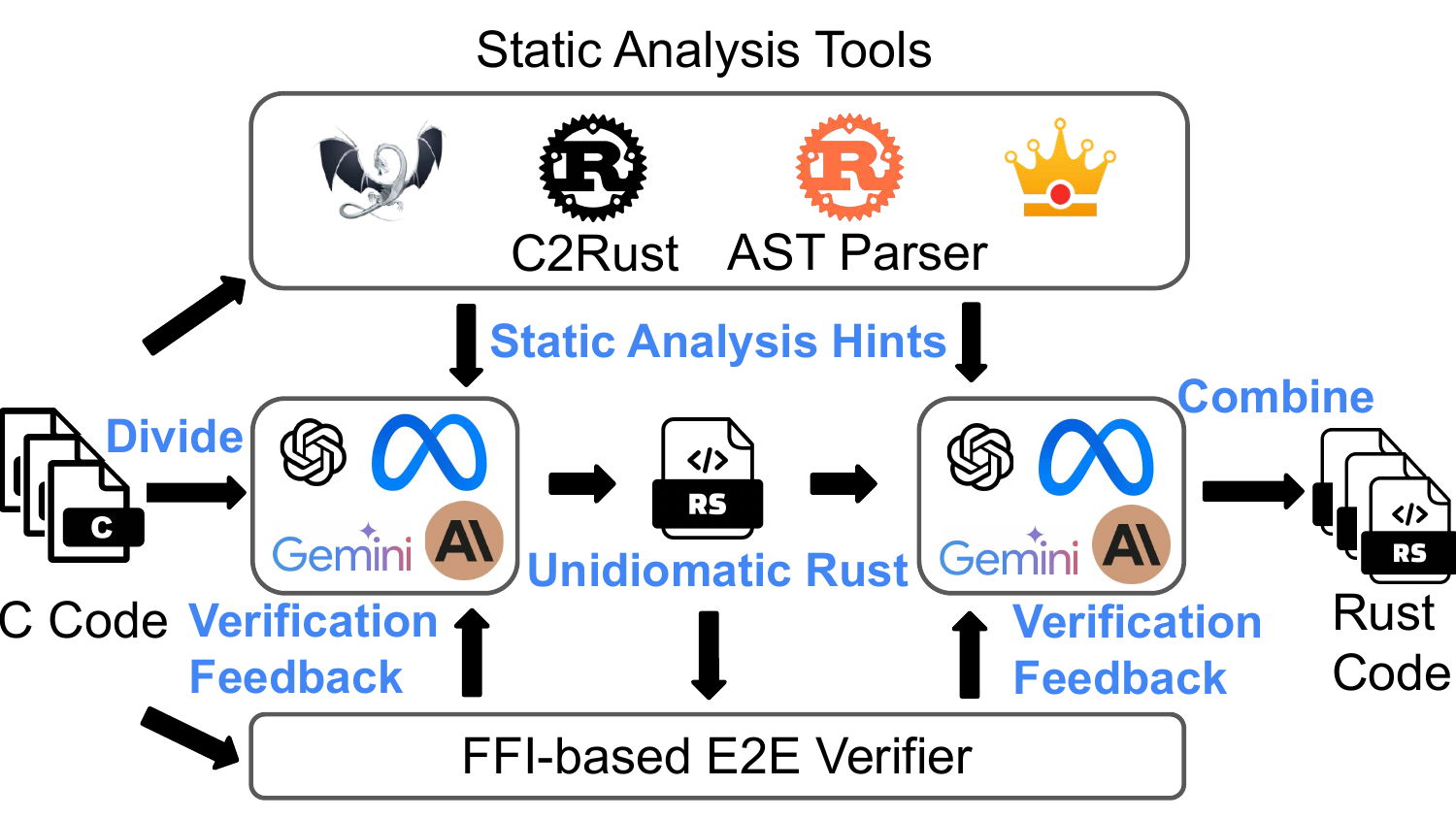}
  \caption{Overview of the \sys{} methodology. \vspace{-5mm}}
  \label{fig:methodology}
\end{figure}

In this paper, we introduce \sys{}, a structure-aware, LLM-driven C-to-Rust translator (Figure~\ref{fig:methodology}). \sys{} follows a two-stage pipeline:
\begin{squishenumerate}
    \item \textbf{C $\to$ Unidiomatic Rust:} Interface-preserving translation that may use \lstinline!unsafe! for low-level operations.
    \item \textbf{Unidiomatic $\to$ Idiomatic Rust:} Behaviorally-equivalent translation that refines to Rust idioms, eliminating \lstinline!unsafe! and migrating C API patterns to Rust equivalents.
\end{squishenumerate}
Static analysis of C code (pointer semantics, dependencies) guides both stages. To verify correctness, we embed the translated Rust with the original C via the Foreign Function Interface (FFI), enabling end-to-end testing on both stages and accept a stage when all end-to-end tests can pass. This decomposition separates syntax from semantics, simplifies the LLM task, and ensures more idiomatic, memory-safe Rust\footnote{\sys{} code is available at \url{https://github.com/qsdrqs/sactor} and datasets are available at \url{https://github.com/qsdrqs/sactor-datasets}}. An example of \sys{} translation process is in Appendix~\ref{app:example}.

\noindent\textbf{LLM orchestration.} \sys{} places the LLM inside a neuro-symbolic feedback loop. Static analysis and a machine-readable interface specification guide prompting; compiler diagnostics and end-to-end tests provide structured feedback. In the idiomatic verification phase, a rule-based harness generator with an LLM fallback completes the feedback loop. This design first ensures semantic correctness in unidiomatic Rust, then refines it into idiomatic Rust, with both stages verifiable in a unified two-step process.

Our contributions are as follows:
\begin{squishitemize}
    \item \textbf{Method:} An LLM-orchestrated, structure-aware two-phase pipeline that separates semantic preservation from idiomatic refinement, guided by static analysis (\S~\ref{sec:method})
    \item \textbf{Verification:} \sys{} verifies both unidiomatic and idiomatic translations via FFI-based testing. During idiomatic verification, it uses a co-produced interface specification to synthesize C/Rust harnesses with an LLM fallback for missing patterns; compiler and test feedback are structured into targeted prompt repairs (\S~\ref{sec:verification}).
    \item \textbf{Evaluation:} Across two datasets (200 programs) and five LLMs, \sys{} reaches 93\% / 84\% end-to-end correctness (DeepSeek-R1) and improves idiomaticity (\S~\ref{subsec:result_measure}). On CRust-Bench (50 samples), unidiomatic translation averages 85\% \emph{function-level success rate} across all samples (82\% aggregated across functions), with 32/50 samples fully translated; idiomatic success is computed on those 32 samples and averages 52\% (43\% aggregated; 8/32 fully idiomatic). On \texttt{libogg} (77 functions), the \emph{function-level success rate} is 100\% for unidiomatic and 53\% and 78\% for idiomatic across GPT-4o and GPT-5, respectively (\S~\ref{subsec:complex}).
    \item \textbf{Diagnostics:} We analyze efficiency, feedback, temperature sensitivity, and failure cases: GPT-4o is the most token-efficient, compilation/testing feedback boosts weaker models by 17\%, temperature has little effect, and reasoning models like DeepSeek-R1 excel on complex bugs such as format-string and array errors (Appendix~\ref{subsub:failure_analysis}).
\end{squishitemize}

%% file: contents/2_background.tex
\section{Background}
\label{sec:background}

\noindent{\bf Primer on C and Rust}: C is a low-level language that provides direct access to memory and hardware through pointers and abstracts machine-level instructions~\citep{tiobe}. While this makes it efficient, it suffers from memory vulnerabilities~\citep{sbufferoverflow, hbufferoverflow,uaf,memoryleak}. Rust, in contrast, provides memory safety without additional performance penalty, {\em and} has the same ability to access low-level hardware as C; it enforces strict compile-time memory safety through {\em ownership}, {\em borrowing}, and {\em lifetimes} to eliminate memory vulnerabilities~\citep{matsakis2014rust, jung2017rustbelt}.

\noindent{\bf Challenges in Code Translation:}
Despite its advantages, and since Rust is relatively new, many widely used system-level programs remain in C.
It is desirable to translate such programs to Rust, but the process is challenging due to fundamental language differences.
Figure~\ref{fig:example} in Appendix~\ref{app:diff_c_rust}
shows an example of a simple C program and its Rust equivalent to illustrate
the differences between two languages in terms of memory management and error handling.
While Rust permits \texttt{unsafe} blocks for C-like pointer operations, their use is discouraged due to the absence of compiler guarantees and their non-idiomatic nature for further maintenance\footnote{Other differences include string representation, pointer usage, array handling, reference lifetimes, and error propagation. A non-exhaustive summary appears in Appendix~\ref{app:diff_c_rust}.}.

%% file: contents/3_related_works.tex
\section{Related Work}
\label{sec:related}

\noindent{\bf LLMs for C-to-Rust Translation:}
Vert~\citep{vert} combines LLM-generated candidates with fuzz testing and symbolic execution to ensure equivalence, but this strict verification struggles with scalability and complex C features. 
Flourine~\citep{flourine} incorporates error feedback and fuzzing, using data type serialization to mitigate mismatches, yet serialization issues still account for nearly half of errors. 
\citet{shiraishi2024context} decompose C programs into sub-tasks (e.g., macros) and translate them with predefined Rust idioms, but evaluate only compilation success without functional correctness. 
\citet{syzygy} employ dynamic analysis to capture runtime behavior as translation guidance, but coverage limits hinder generalization across execution paths. 
\citet{c2saferrust} refine C2Rust outputs with LLMs to reduce unidiomatic constructs (\texttt{unsafe}, \texttt{libc}), but remain constrained by C2Rust's preprocessing, which strips comments and directives~(\S~\ref{sec:unidiomatic_rust_translation}) and reduces context for idiomatic translation.

\noindent{\bf Non-LLM Approaches for C-to-Rust Translation:}
C2Rust~\citep{c2rust} translates by converting C ASTs into Rust ASTs and applying rule-based transformations. While syntactically correct, the results are structural translations that rely heavily on \texttt{unsafe} blocks and explicit type conversions, yielding low readability. 
Crown~\citep{crown} introduces static ownership tracking to reduce pointer usage in generated Rust code. 
\citet{hong2024don} focus on handling return values in translation, while \citet{ling2022rust} rely on rules and heuristics. 
Although these methods reduce some \texttt{unsafe} usage compared to C2Rust, the resulting code remains largely unidiomatic.

%% file: contents/5_method_details.tex
\section{\sys{} Methodology}
\label{sec:method}

We propose \sys{},  an LLM-driven C-to-Rust translation tool using a two-step
translation methodology.
As Rust and C differ substantially in semantics (\S~\ref{sec:background}), \sys{} augments the LLM with static-analysis-derived ``hints'' that capture semantic information in the C code. The four main stages of \sys{} are outlined below.

\subsection{Task Division}
\label{sec:task_division}

We begin by dividing the program into smaller parts that can be processed by the LLM independently. This enables the LLM to focus on a narrower scope for each translation task and ensures the program fits within its context window.
This strategy is supported by studies showing that LLM performance degrades on long-context understanding and generation tasks~\citep{liu2024longgenbench, li2024long}.
By breaking the program into smaller pieces, we can mitigate these limitations and improve performance on each individual task.
To facilitate task division and extract relevant language information -- such as definitions, declarations, and dependencies -- from C code, we developed a \textit{static analysis tool} called \texttt{C Parser} based on \texttt{libclang} (a library that provides a C compiler interface, allowing access to semantic information of the code). 

Our \texttt{C Parser} analyzes the input program and splits the program into fragments consisting of a single type, global variable, or function definition. This step also extracts semantic dependencies between each part (e.g., a function definition depending on a prior type definition). We then process each program fragment in dependency order: all dependencies of a code fragment are processed before the fragment.
Concretely, \texttt{C Parser} constructs a directed dependency graph whose nodes are types, global variables, and functions, and whose edges point from each item to the items it directly depends on. We compute a translation order by repeatedly selecting items whose dependencies have already been processed. If the dependency graph contains a cycle, \sys{} currently treats this as an unsupported case and terminates with an explicit error.
In addition, to support real-world C projects, \sys makes use of the C project compile commands generated by the \texttt{make} tool and performs preprocessing on the C source files.
In Appendix~\ref{app:task_division}, we provide more details on how we preprocess source files and divide  programs.

\subsection{Translation}\label{sec:translation}

To ensure that each program fragment is translated only {\em after} its dependencies have been processed, we begin by translating data types, as they form the foundational elements for functions. This is followed by global variables and functions. %
We divide the translation process into two steps.

\noindent \underline{\em Step 1. Unidiomatic Rust Translation:}
\label{sec:unidiomatic_rust_translation}
We aim to produce interface equivalent Rust code from the original C code, which allows the use of \texttt{unsafe} blocks to do pointer manipulations and C standard library functions while keeping the same interface as original C code. For data type translation, we leverage information from C2Rust~\citep{c2rust} to help the conversion. While C2Rust provides reliable data type translation, {\em it struggles with function translation} due to its compiler-based approach, which omits source-level details like comments, macros, and other elements. These omissions significantly reduce the readability and usability of the generated Rust code. Thus, we use C2Rust only for data type translation, and use an LLM to translate global variables and functions. For functions, we rely on our \texttt{C Parser} to automatically extract dependencies (e.g., function signatures, data types, and global variables) and reference the corresponding Rust code. This approach guides the LLM to accurately translate functions by leveraging the previously translated components and directly reusing or invoking them as needed.

\noindent \underline{\em Step 2. Idiomatic Rust Translation:}
The goal of this step is to refine unidiomatic Rust into idiomatic Rust by removing \texttt{unsafe} blocks and following Rust idioms.
This stage focuses on rewriting behavioral-equivalent but low-level constructs into type-safe abstractions while preserving behavior verified in the previous step.
Handling pointers from C code is a key challenge, as they are considered unsafe in Rust. Unsafe pointers should be replaced with Rust types such as references, arrays, or owned types. To address this, we use Crown~\citep{crown} to facilitate the translation by analyzing pointer {mutability}, {fatness} (e.g., arrays), and {ownership}. This information provided by Crown helps the LLM assign appropriate Rust types to pointers. Owned pointers are translated to \texttt{Box}, while borrowed pointers use references or smart pointers. Crown assists in translating data types like \texttt{struct} and \texttt{union}, which are processed first as they are often dependencies for functions. For function translations, Crown analyzes parameters and return pointers, while local variable pointers are inferred by the LLM. Dependencies are extracted using our {\texttt{C Parser}} to guide accurate function translation.
The idiomatic code is produced together with an interface transformation specification, forms the input to the verification step in \S~\ref{sec:verification}.

\subsection{Verification}
\label{sec:verification}

To verify the equivalence between source and target languages, prior work has relied on symbolic execution and fuzz testing, are impractical for real-world C-to-Rust translation (details in Appendix~\ref{app:testing}).
We instead validate correctness through \textit{soft equivalence}: ensuring functional equivalence of the entire program via end-to-end (E2E) tests.
This avoids the complexity of generating specific inputs or constraints for individual functions and is well-suited for real-world programs where such E2E tests are often available and reusable.
Correctness confidence in this framework depends on the code coverage of the E2E tests: the broader the coverage, stronger the assurance of equivalence.

\noindent \underline{\em Verifying Unidiomatic Rust Code.}
This is straightforward, as it is semantically equivalent to the original C code and maintains compatible function signatures and data types, which ensures a consistent Application Binary Interface (ABI) between the two languages and enabling direct use of the FFI for cross-language linking.
The verification process involves two main steps:
First, the unidiomatic Rust code is compiled using the Rust compiler to check for successful compilation. Then, the original C code is recompiled with the Rust translation linked as a shared library. This setup ensures that when the C code calls the target function, it invokes the Rust translation instead. To verify correctness, {\em E2E tests are run on the entire program}, comparing the outputs of the original C code and the unidiomatic Rust translation. If all tests pass, the target function is considered verified.

\noindent \underline{\em Verifying Idiomatic Rust Code.}
Idiomatic Rust diverges from the original C program in both types and function
signatures, producing an \emph{ABI mismatch} that prevents direct linking into
the C build. We therefore verify it via a {\em synthesized, C-compatible test harness together with
E2E tests}.

During idiomatic translation, \sys co-produces a small, machine-readable {\em specification} (SPEC) for each function/struct. The SPEC captures, in a compact form, how C-facing values map to idiomatic Rust, including the expected pointer shape (\texttt{slice}/\texttt{cstring}/\texttt{ref}), where lengths come from (a sibling field or a constant), and basic nullability and return conventions; it also allows marking fields that should be compared in self-checks. A rule-based generator consumes the SPEC to synthesize a C-compatible harness that bridges from the C ABI to idiomatic code and backwards. Figure~\ref{fig:harness_gen} shows the schematic, and Table~\ref{tab:spec_rules} summarizes current supported patterns; Appendix~\ref{app:spec_rules} presents a detailed exposition of the SPEC-driven harness generation technique (rules and design choices), and Appendix~\ref{app:test_harness_example} provides a concrete example of the generated harness.
For structs, the SPEC defines bidirectional converters between the C-facing and idiomatic layouts, validated by a lightweight roundtrip test that checks the fields marked as comparable for consistency after conversion. When the SPEC includes a pattern the generator does not yet implement (e.g., aliasing/offset views or unsupported pointer kinds or types), we emit a localized TODO and use an LLM guided by the SPEC to fill only the missing conversions. Finally, we compile the idiomatic crate and the generated harness, link them into the original C build via FFI, and run the program's existing E2E tests; passing tests validate the idiomatic translation under the coverage of those tests, while failures trigger the feedback procedure in \S~\ref{subsec:feedback}.

\noindent \underline{\em Feedback Mechanism.}\label{subsec:feedback}
For failures, we feed structured signals back to translation: compiler errors guide fixes for build breaks; for E2E failures we use the Rust procedural macro to automatically instrument the target to log salient inputs/outputs, re-run tests, and return the traces to the translator for refinement.

\subsection{Code Combination}
\label{sec:code_combination}

By translating and verifying all functions and data types, we integrate them
into a unified Rust codebase. We first collect the translated Rust code from
each subtask and remove duplicate definitions and other redundancies required
only for standalone compilation. The cleaned code is then organized into a
well-structured Rust implementation of the original C program. Finally, we run
end-to-end tests on the combined program to verify the correctness of the final
Rust output. If all tests pass, the translation is considered successful.

%% file: contents/6_experimental_setup.tex
\section{Experimental Setup}
\label{sec:setup}

\subsection{Datasets Used}

For the selection of datasets for evaluation, we consider the following criteria:
\vspace{-2mm}
\begin{myitemize}
\itemsep0em
\item {\em Sufficient Number:} The dataset should contain a substantial number of C programs to ensure a robust evaluation of the approach's performance across a diverse set of examples.
\item {\em Presence of Non-Trivial C Features:} The dataset should include C programs with advanced features such as multiple functions, \texttt{struct}s, and other non-trivial constructs as it enables the evaluation to assess the approach's ability to handle complex features of C.
\item {\em Availability of E2E Tests:} The dataset should either include E2E tests or make it easy to generate them as this is essential for accurately evaluating the correctness of the translated code.
\end{myitemize}
\vspace{-2mm}

Based on the above criteria, we evaluate on two widely used program suites in the translation literature: TransCoder-IR~\citep{transcoderir} and Project CodeNet~\citep{codenet}. Complete details for these datasets are in Appendix~\ref{app:datasets}.
For TransCoder-IR and CodeNet, we randomly sample 100 C programs from each (for CodeNet, among programs with external inputs) to ensure computational feasibility while maintaining statistical significance.

To better reflect the language features of real-world C codebases and allow test reuse (\S~\ref{subsec:complex}), we also evaluate on two targets: (i) a 50-sample subset of CRust-Bench \citep{khatry2025crust} and (ii) the \texttt{libogg} multimedia container library \citep{libogg}. In CRust-Bench, we exclude entries outside our pipeline's scope (e.g., circular dependencies or compiler-specific intrinsics). \texttt{libogg} is a real-world C project of about 2,000 lines of code with 77 functions involving non-trivial \texttt{struct}s, \texttt{buffer}s, and pointer manipulation. Both benchmarks reuse their upstream end-to-end tests to verify the translated code.

\subsection{Evaluation Metrics}
\label{subsec:metrics}

\noindent{\bf Success Rate:} This is defined as the ratio of the number of programs that can (a) successfully be translated to Rust, and (b) successfully pass the E2E tests for both unidiomatic and idiomatic translation phases to the total number of programs. To enable the LLMs to utilize feedback from previous failed attempts, we allow the LLM to make up to 6 attempts for each translation process.

\noindent{\bf Idiomaticity:} To evaluate the idiomaticity of the translated code, we use three metrics:

\vspace{-2mm}
\begin{myitemize}
\itemsep0em
\item {\em Lint Alert Count} is measured by running Rust-Clippy~\citep{clippy}, a tool that provides lints on unidiomatic Rust (including improper use of \texttt{unsafe} code and other common style issues). By collecting the warnings and errors generated by {Rust-Clippy} for the translated code, we can assess its idiomaticity: fewer alerts indicate more idiomaticity. Previous translation works~\citep{vert, flourine} have also used {Rust-Clippy}. %

\item {\em Unsafe Code Fraction}, inspired by~\citet{shiraishi2024context}, is defined as the ratio of tokens inside \texttt{unsafe} code blocks or functions to total tokens for a single program. High usage of \texttt{unsafe} is considered unidiomatic, as it bypasses compiler safety checks, introduces potential memory safety issues and reduces code readability.

\item {\em Unsafe Free Fraction} indicates the percentage of translated programs in
a dataset that do not contain any \texttt{unsafe} code. Since \texttt{unsafe} code represents potential points where the compiler cannot guarantee safety, this metric helps determine the fraction of results that can be achieved without relying on \texttt{unsafe} code.
\end{myitemize}
\vspace{-2mm}

\subsection{LLMs Used}
\label{subsec:models}

We evaluate 6 models across different experiments. On the two datasets
(TransCoder-IR and CodeNet) we use four non-reasoning models\textemdash GPT-4o
(OpenAI), Claude 3.5 Sonnet (Anthropic), Gemini 2.0 Flash (Google), and Llama
3.3 70B Instruct (Meta), and one reasoning model DeepSeek-R1 (DeepSeek). For
real-world codebases, we run GPT-4o on CRust-Bench and run both GPT-4o and GPT-5 on \texttt{libogg}.
Model configurations appear in Appendix~\ref{app:llm_config}.

%% file: contents/7_evaluation.tex
\section{Evaluation}
\label{sec:results}

Through our evaluation, we answer:
(1) How successful is \sys{} in generating idiomatic Rust code using different LLM models?;
(2) How idiomatic is the Rust code produced by \sys{} compared to existing approaches?; and
(3) How well does \sys{} generalize to real-world C codebases?

Our results show that: (1) DeepSeek-R1 achieves the highest success rates (93\%) with \sys{}
on TransCoder-IR and also reaches the highest success rates (84\%) on
Project CodeNet (\S~\ref{subsub:success_rate}), while failure reasons vary between datasets
and models (Appendix~\ref{subsub:failure_analysis});
(2) \sys{}'s idiomatic translation results outperforms all previous baselines, producing Rust code with fewer Clippy warnings and 100\% unsafe-free translations (\S~\ref{subsec:result_measure}); and %
(3) For real-world codebases (\S~\ref{subsec:complex}), \sys{} attains strong
unidiomatic success and moderate idiomatic success: on CRust-Bench, unidiomatic
averages 85\% across 50 samples
(82\% aggregated across 966 functions; 32/50 fully translated) and idiomatic averages 52\% across 32
samples that fully translated into unidiomatic Rust (43\% aggregated across 580 functions; 8/32 fully translated); on \texttt{libogg}
unidiomatic reaches 100\% and idiomatic spans 53\% and 78\% for GPT-4o and GPT-5, respectively.
Failures concentrate at ABI/type boundaries and harness
synthesis (pointer/slice shape, length sources, lifetime or mutability), with
additional cases from unsupported features and borrow/ownership pitfalls.
Overall, improving the model itself alleviates a subset of failure modes; for a fixed model, strengthening the framework and interface rules also improves outcomes but remains limited when confronted with previously unseen patterns.

We also evaluate the computational cost of \sys{} (Appendix~\ref{subsec:cost}), the impact of the feedback mechanism
(Appendix~\ref{app:ablation}),
and temperature settings (Appendix~\ref{app:temperature})
. GPT-4o and Gemini 2.0 achieve the best cost-performance balance, while Llama
3.3 consumes the most tokens among non-reasoning models. DeepSeek-R1 uses 3-7 $\times$ more tokens than others. The feedback mechanism boosts Llama 3.3's
success rate by 17\%, but has little effect on GPT-4o, suggesting it benefits
lower-performing models more. Temperature has minimal impact.

\input{contents/7.1_idiomatic}

\input{contents/7.3_prior}
\input{contents/7.4_complex}

%% file: contents/7.1_idiomatic.tex
\subsection{Success Rate Evaluation}
\label{subsub:success_rate}

\begin{figure}[ht]
    \vspace{-2mm}
    \centering
    \includegraphics[width=0.7\linewidth]{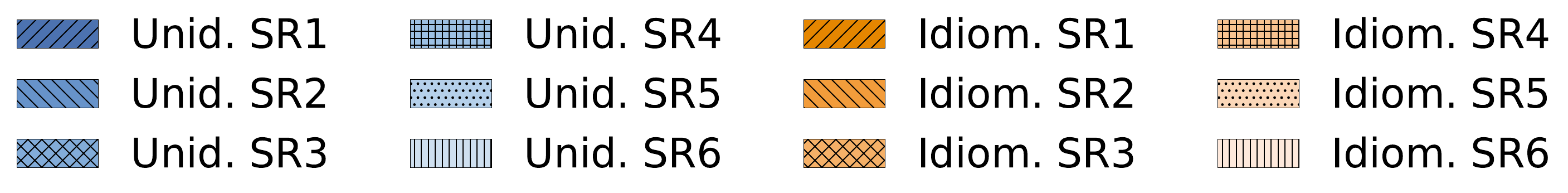}
    \begin{subfigure}{0.4\linewidth}
        \centering
        \includegraphics[width=\linewidth]{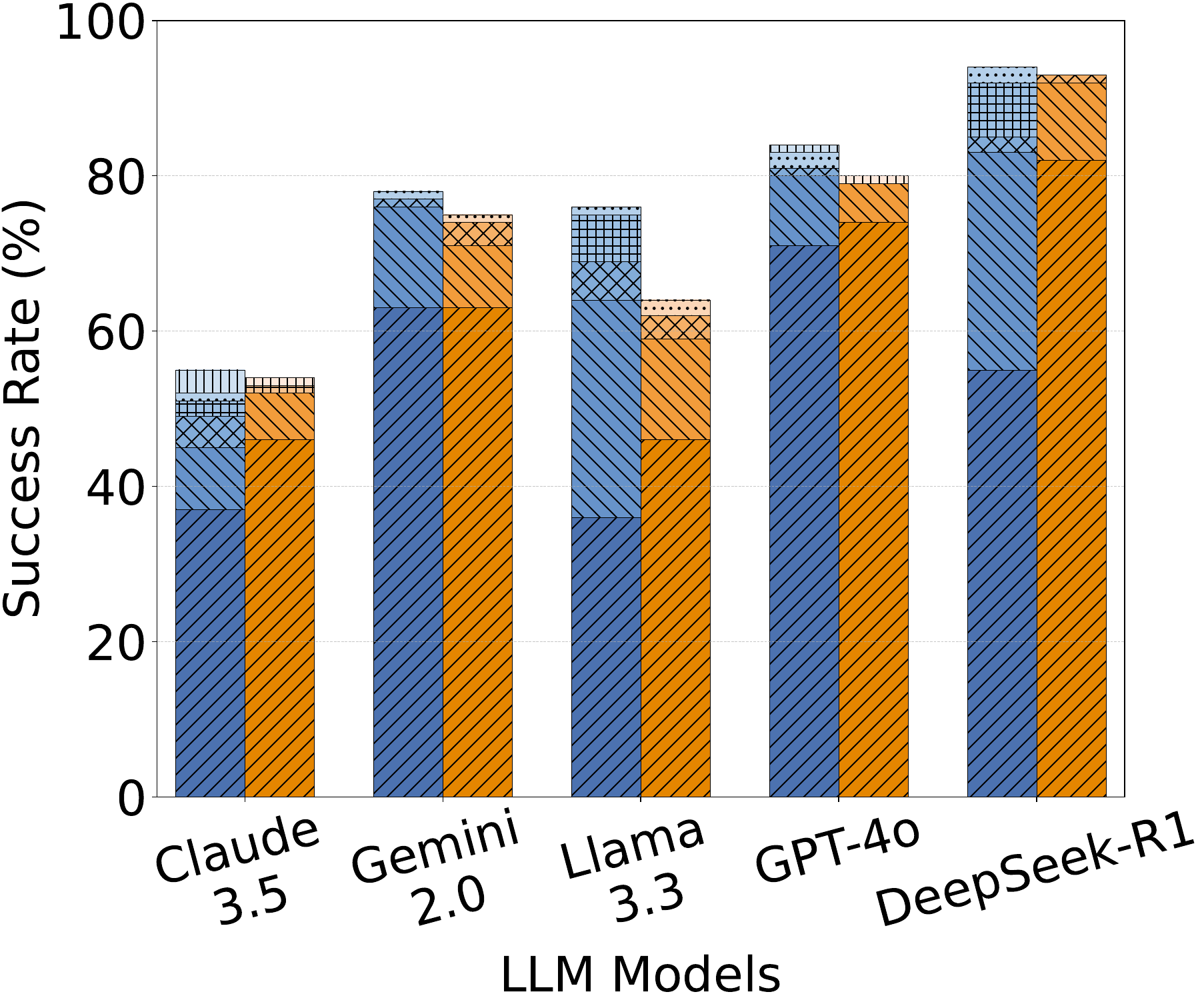}
        \caption{TransCoder-IR SR}
        \label{fig:success_rate_transcoder}
    \end{subfigure}
    \hspace{-2mm}
    \begin{subfigure}{0.4\linewidth}
        \centering
        \includegraphics[width=\linewidth]{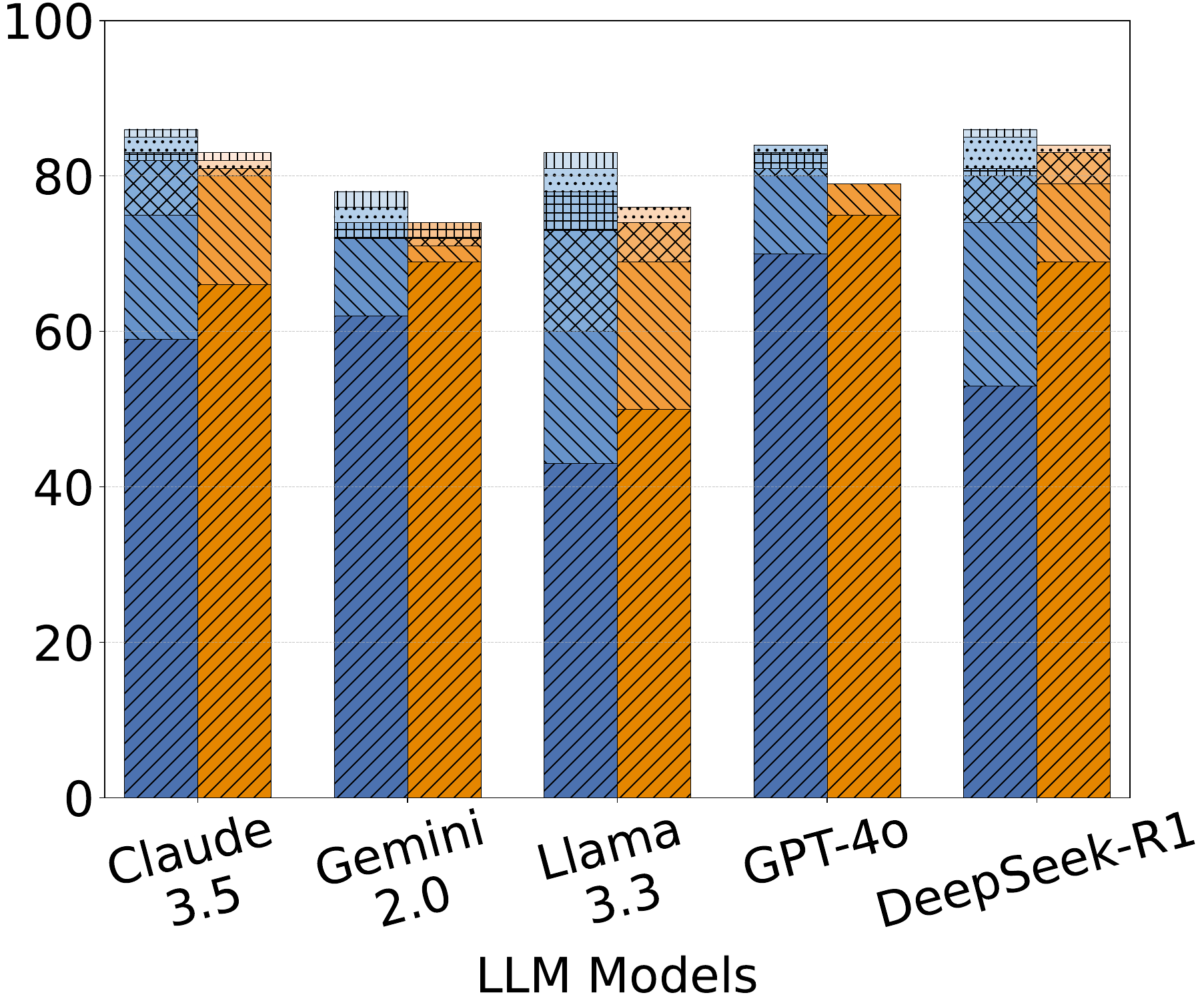}
        \caption{CodeNet SR}
        \label{fig:success_rate_codenet}
    \end{subfigure}
    \caption{Success rates (SR) across different LLM models for the
    TransCoder-IR and CodeNet datasets. SR 1-6 represent the number of attempts
    made to achieve a successful translation. Unid. and Idiom. denote unidiomatic
    and idiomatic translation steps, respectively.}
    \label{fig:success_rates}
\end{figure}

We evaluate the success rate (as defined in \S~\ref{subsec:metrics}) for the two datasets on different models. For idiomatic translation, we also plot how many attempts are needed.

\noindent{\bf (1) TransCoder-IR} (Figure~\ref{fig:success_rate_transcoder}):
DeepSeek-R1 achieves the highest success rate (SR) in both unidiomatic (94\%)
and idiomatic (93\%) steps, only 1\% drops in the idiomatic translation step,
demonstrating strong consistency in code translation.
GPT-4o follows with 84\% in the unidiomatic step and 80\% in the idiomatic step.
Gemini 2.0 comes next with 78\% and 75\%, respectively. Claude 3.5 struggles in
the unidiomatic step (55\%) but does not show substantial degradation when
converting unidiomatic Rust to idiomatic Rust (54\%, only a 1\% drop), but
it is still the worst model compared to the others. Llama 3.3
performs well in the unidiomatic step (76\%) but drops significantly in the
idiomatic step (64\%), and requiring more attempts for correctness.

\noindent{\bf (2) Project CodeNet} (Figure~\ref{fig:success_rate_codenet}):
DeepSeek-R1 again leads with 86\% in the unidiomatic step and 84\% in the
idiomatic step, showing only a 2\% drop in the idiomatic translation step.
Claude 3.5 follows closely with 86\% success rate in the unidiomatic step and 83\% in the
idiomatic step. GPT-4o performs consistently well in the unidiomatic step (84\%) but drops to 79\% in the idiomatic step,
indicating a 5\% drop between the two steps.
Gemini 2.0 follows with 78\% in the unidiomatic step and 74\% in the idiomatic step, showing consistent performance
between two datasets. Llama 3.3 still exhibits significant drops (83\% to 76\%) in both steps
and finishes last in the idiomatic step.

The results demonstrates that DeepSeek-R1's SRs remain high and consistent--94\%/93\%
(unidiomatic/idiomatic) on TransCoder-IR versus 86\%/84\% on CodeNet--while
other models exhibit notable performance drops when moving to TransCoder-IR.
This suggests that models with reasoning capabilities may be better for handling complex code logic and
data manipulation.

%% file: contents/7.3_prior.tex
\subsection{Measuring Idiomaticity}
\label{subsec:result_measure}

We compare our approach with four baselines: C2Rust~\citep{c2rust},
Crown~\citep{crown}, C2SaferRust~\citep{c2saferrust} and Vert~\citep{vert}. Of
these baselines, C2Rust is the most versatile\footnote{Versatility refers to an approach's applicability to diverse C programs.}, supporting most C programs,
while Crown is also broad but lacks support for some language features.
C2SaferRust focuses on refining the unsafe code produced by C2Rust, allowing it to handle a wide range of C programs. In contrast, Vert targets a specific subset of simpler C programs.
We assess the idiomaticity of Rust code generated by C2Rust, Crown, and C2SaferRust on both datasets. Since Vert produced Rust code only for TransCoder-IR, we evaluate it solely on this dataset.
All the experiments are conducted using GPT-4o as the LLM for baselines and our approach, with max 6 attempts per translation.

\afterpage{
\begin{figure}[tbhp]
    \centering
    \begin{minipage}[b]{.49\textwidth}
    \centering
    \includegraphics[width=0.9\textwidth]{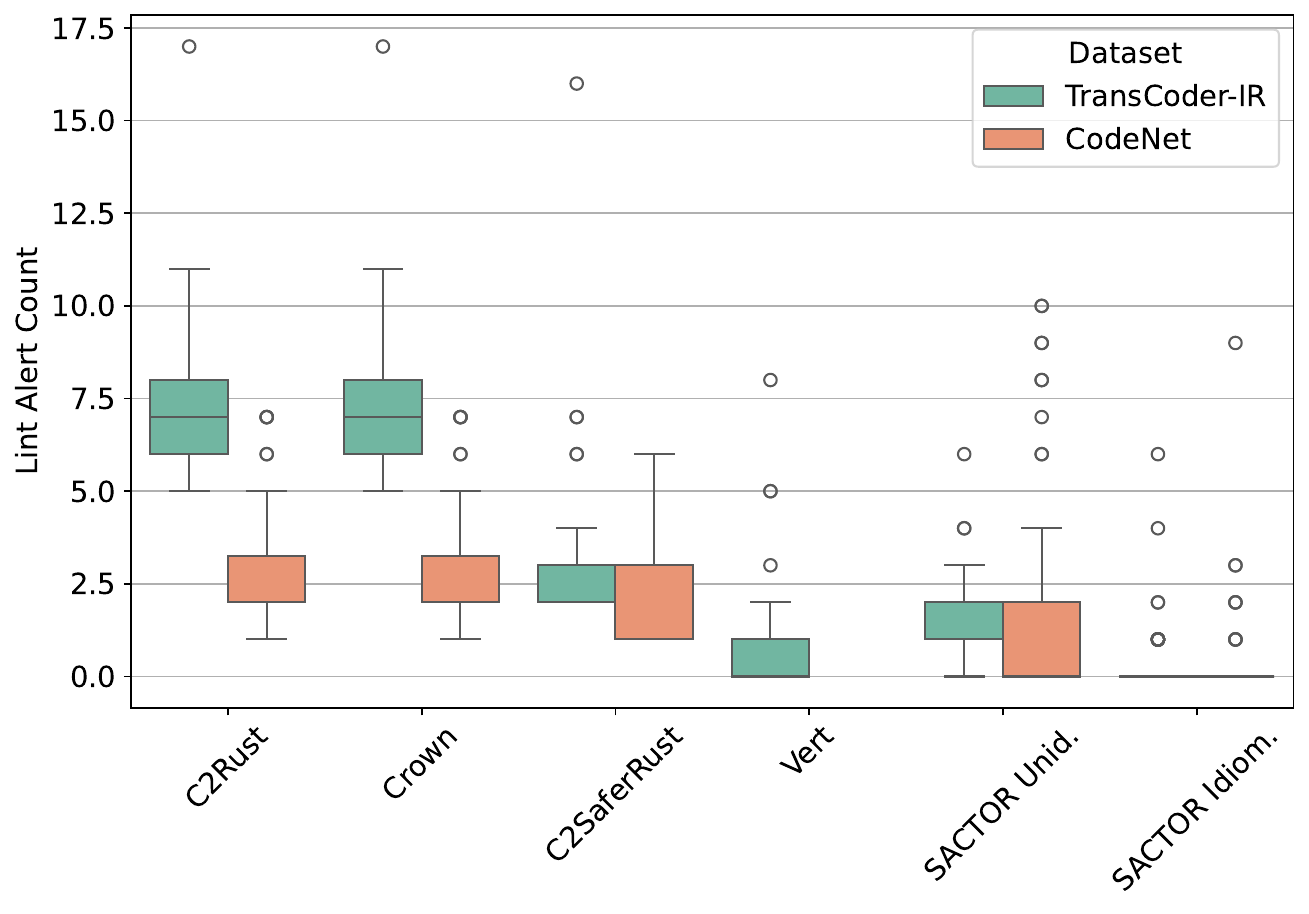}
    \vspace{-2mm}
    \caption{Total clippy issues (warnings + errors) across different method.}
    \label{fig:idiomaticity}
    \end{minipage}
    \hfill
    \begin{minipage}[b]{.49\textwidth}
    \centering
    \scriptsize
    \begin{adjustbox}{max width=\columnwidth}
    \begin{tabular}{lcccc}
        \toprule
        \textsc{Method} & \textsc{Dataset} & \textsc{SR (\%)} & \textsc{UF (\%)} & \textsc{AU (\%)} \\
        \midrule
        C2Rust          & TransCoder-IR    & 100              & 0                & 100              \\
                        & CodeNet          & 100              & 0                & 75.9             \\
        \midrule
        Crown           & TransCoder-IR    & 100              & 0                & 100              \\
                        & CodeNet          & 100              & 0                & 75.9             \\
        \midrule
        C2SaferRust     & TransCoder-IR    & 90               & 45.6             & 10.8             \\
                        & CodeNet          & 93               & 0                & 75.8             \\
        \midrule
        Vert            & TransCoder-IR    & 92               & 95.7             & 1.6              \\
        \midrule
        \sys{} (Unid.)  & TransCoder-IR    & 84               & 3.6              & 91.7             \\
                        & CodeNet          & 84               & 1.1              & 42.7             \\
        \midrule
        \sys{} (Idiom.) & TransCoder-IR    & 80               & \textbf{100}     & \textbf{0}       \\
                        & CodeNet          & 79               & \textbf{100}     & \textbf{0}       \\
        \bottomrule
    \end{tabular}
    \end{adjustbox}
    \captionof{table}{Unsafe code statistics. UF denotes ``Unsafe Free" and AU denotes ``Avg. Unsafe."}
    \label{tab:unsafe_stats}
    \end{minipage}
\end{figure}
}

\noindent{\bf Results:} Figure~\ref{fig:idiomaticity} presents the lint alert count (sum up of Clippy warnings and errors count for a single program) across all approaches. C2Rust consistently exhibits high Clippy issues, and Crown shows little improvement over C2Rust, indicating both struggle to generate idiomatic Rust.
C2SaferRust reduces Clippy issues, but it still retains a significant number of
warnings and errors.
Notably, even the unidiomatic output of \sys{} surpasses all of these 3.
This underscores the advantage of LLMs over rule-based methods.
While Vert improves idiomaticity, \sys{}'s idiomatic phase yields fewer Clippy issues, outperforming some existing LLM-based approaches.

Table~\ref{tab:unsafe_stats} summarizes unsafe code statistics. Unsafe-Free indicates the percentage of programs without unsafe code, while Avg. Unsafe represents the average proportion of unsafe code across all translations. %
C2Rust and Crown generate unsafe code in all programs with a high average unsafe
percentage. C2SaferRust has the ability to reduce unsafe code and able to
generate unsafe-free programs in some cases (45.6\% in TransCoder-IR), but
cannot sufficiently reduce the unsafe uses in the CodeNet dataset.
Vert has a higher success rate than \sys{} but occasionally introduces unsafe code. \sys{}'s unidiomatic phase retains C semantics, leading to a high unsafe percentage. However, its idiomatic phase eliminates all unsafe code, achieving a 100\% Unsafe-Free rate.

%% file: contents/7.4_complex.tex
\subsection{Real-world Code-bases}
\label{subsec:complex}

To evaluate \sys{}'s performance on two real-world code-bases, we run the translation process up to three times per sample, with \sys attempts to translate each function, struct and global variable at most six attempts in each run.
For \texttt{libogg}, we also experiment with both GPT-4o and GPT-5 to compare their performance.

\input{contents/7.4_crust_bench}

\input{contents/7.4_libogg}

%% file: contents/7.4_crust_bench.tex
\paragraph{CRust-Bench.}
Measured at the function level, the
mean per-sample translation success rate is 85.15\%. Aggregated across the 50 samples,
\sys{} translates 788 of 966 functions (81.57\% combined). 32 samples
achieve 100\% function-level translation, i.e., the entire C codebase for the
sample is translated to unidiomatic Rust.
For idiomatic translation, we evaluate only
on the 32 samples whose unidiomatic stage reached 100\% function-level
translation. On these samples, the mean per-sample function translation rate is
51.85\%. Aggregated across them, \sys{} translates 249 of 580 functions (42.93\%
combined); 8 samples achieve 100\% function-level idiomatic translation, which the
entire C codebases are translated to idiomatic Rust.

\begin{table*}[t]
\centering
\scriptsize
\begin{tabular}{lccccc}
\toprule
\textsc{Stage} & \textsc{Samples Eval.} & \textsc{Per-sample SR (func.)} & \textsc{Aggregated SR (func.)} & \textsc{Full SR} & \textsc{Avg. Lint / Function} \\
\midrule
Unidi. & 50 & 85.15\% & 788 / 966 (81.57\%) & 32 / 50 (64.00\%) & 2.96 \\
Idiom. & 32\textsuperscript{$\dagger$} & 51.85\% & 249 / 580 (42.93\%) & 8 / 32 (25.00\%) & 0.28 \\
\bottomrule
\end{tabular}
\caption{CRust-Bench function-level translation results. Success rate (SR) is averaged per-sample; $\dagger$ idiomatic stage is evaluated only on samples whose unidiomatic pass fully translated all functions.}
\label{tab:crust_results}
\end{table*}

\noindent Table~\ref{tab:crust_results} summarizes stage-level outcomes.

\noindent\underline{\textit{Observations and failure modes.}} We organize failures into
five main categories.
\emph{(1) Interface/name drift:} Symbol casing or exact-name mismatches (e.g., \texttt{CamelCase} vs. \texttt{snake\_case}).
\emph{(2) Semantic mapping errors:} Mistakes in translating C constructs to idiomatic Rust (e.g., pointer-of-pointer vs. \texttt{Vec}, shape drift, lifetime or mutability issues).
\emph{(3) C-specific features:} Incomplete handling some features like function pointers and C variadics.
\emph{(4) Borrowing and resource-model violations:} Compile-time borrow-checker errors in idiomatic Rust bodies (e.g., overlapping borrows in updates).
\emph{(5) Harness/runtime faults:} Faulty test harnesses translation (e.g. buffer mis-sizing, out-of-bounds access).
Other minor cases include \emph{unsupported intrinsics} (SIMD) and \emph{global-state divergence} (shadowed globals).
Table~\ref{tab:crust_failures} (in Appendix~\ref{app:crust_failures}) summarizes each sample's outcome and its primary cause.

\noindent\underline{\textit{Idiomaticity.}} Unidiomatic outputs exhibit many lint alerts and
heavy reliance on \texttt{unsafe}: the mean Clippy alert sum is 50.14 per
sample (2.96 per function); the mean unsafe fraction is 97.86\% with an
unsafe-free rate of 0\%. Idiomatic outputs reverse this profile: the mean Clippy
alert sum drops to only 2.27 per sample (0.28 per function); the mean unsafe
fraction is 0\% with a 100\% unsafe-free rate.

%% file: contents/7.4_libogg.tex
\paragraph{Libogg.}

\input{figures/libogg_table}

The unidiomatic and idiomatic translations of all structs and global variables are successful with each LLM model. For functions, the result is summarized in Table~\ref{tab:libogg-eval}.
\sys{} succeeds in all functions' unidiomatic translations. For idiomatic translations, \sys{}'s success rate is 53\% and \sys{} takes 2.00 attempts on average to produce a correct translation with GPT-4o. For GPT-5, the performance is significantly better with a success rate of 78\% and average number of attempts of 1.25.

\noindent\underline{\textit{Observations and failure modes.}} The most significant reasons for failed idiomatic translations include: 
\textit{(1)} failure to pass tests due to mistakes in translating pointer manipulation and heap memory management; 
\textit{(2)} compile errors in translated functions, especially arising from violation of Rust safety rules on lifetimes, borrowing and mutability; 
\textit{(3)} failure to generate compilable test harnesses for data types with pointers and arrays. 
GPT-5 performs significantly better than GPT-4o. For example, GPT-5 only have one failure caused by a compile error in the translated function, in contrast to six compile error failures with GPT-4o, which shows the progress 
of GPT-5 in understanding Rust grammar and fixing compile errors. More details can be found in Appendix~\ref{app:libogg-outcome}.

\noindent\underline{\textit{Idiomaticity.}} \sys's unidiomatic translations cause lint alerts largely due to the use of \texttt{unsafe} code while idiomatic translations lead to very few lint alerts, \textit{i.e.}, fewer than 0.3 alerts per function on average (Table~\ref{tab:libogg-eval}).
With each model, the unidiomatic translations are all in \texttt{unsafe} code but the idiomatic translations are all in \texttt{safe} code. As a result, the idiomatic translations have an avg. unsafe fraction of 0\% and unsafe-free fraction of 100\%. The unidiomatic translations are the opposite.

%% file: figures/libogg_table.tex
\begin{table}[tbp]
\centering
\scriptsize
\begin{adjustbox}{max width=\columnwidth}
\begin{tabular}{lccc}
\toprule
 \textsc{Step (model)  }   &       \textsc{SR (\%)} & \textsc{Avg. lint / Function} & \textsc{Avg.  attempt}  \\
 \midrule
Unid. (GPT-4o)  &   100  &           1.45         &   1.52             \\
Idiom. (GPT-4o)    &   53   &          0.28          &  2.00           \\
\midrule
Unid. (GPT-5)   &  100    &           1.45           &   1.04             \\
Idiom. (GPT-5)     &   78  &           0.23               &   1.25            \\
\bottomrule
\end{tabular}
\end{adjustbox}
\caption{Evaluation of \sys{}'s function translation on libogg. ``Unid.''/``Idiom.'' denotes unidiomatic/idiomatic translation. ``SR'' is the success rate of translating functions. ``Avg. lint''/``Avg. attempt'' is the average lint alert count/average number of attempts, for functions that both LLM models succeed in translating. \vspace{-5mm}}
\label{tab:libogg-eval}
\end{table}

%% file: contents/8_conclusion.tex
\section{Conclusions}
\label{sec:conclusion}

Translating C to Rust enhances memory safety but remains error-prone and often unidiomatic. 
While LLMs improve translation, they still lack correctness guarantees and struggle with semantic gaps. 
\sys{} addresses these through a two-stage pipeline: preserving ABI interface first, then refining to idiomatic Rust.
Guided by static analysis and validated via FFI-based testing, \sys{} achieves high correctness and idiomaticity across multiple benchmarks, surpassing prior tools.
Remaining challenges include stronger correctness assurance, richer C-feature coverage, and improved scalability and efficiency (see \S\ref{sec:limitation}). 
Example prompts appear in Appendix~\ref{app:prompts}.

%% file: contents/limitations.tex
\section{Limitations}
\label{sec:limitation}

While \sys{} is effective in producing correct, idiomatic Rust, several limitations remain:
\begin{squishitemize}
  \item \textbf{Test coverage dependence.} Our soft-equivalence checks rely on existing end-to-end tests; shallow or incomplete coverage can miss subtle semantic errors. Integrating fuzzing or test generation could raise coverage and catch corner cases.
  \item \textbf{Model variance.} Translation quality depends on the underlying LLM. Although GPT-4o and DeepSeek-R1 perform well in our study, other models show lower accuracy and stability.
  \item \textbf{Unsupported C features.} Complex macros, pervasive function pointers, global state, C variadics and inline assembly are only partially handled, limiting applicability to such codebases (see \S~\ref{subsec:complex}).
  \item \textbf{Static analysis precision.} Current analysis may under-specify aliasing, ownership, and pointer shapes in challenging code, leading to adapter/spec errors. Stronger analyses could improve mapping and reduce retries.
  \item \textbf{Harness generation stability.} The rule-based generator with LLM fallback can still emit incomplete or brittle adapters on complex patterns (e.g., unusual pointer shapes or length expressions), causing otherwise-correct translations to fail verification. Hardening rules and reducing reliance on the fallback should improve robustness and reproducibility.
  \item \textbf{Cost and latency.} Multi-stage prompting, compilation, and test loops incur non-trivial token and time costs, which matter for large-scale migrations.
\end{squishitemize}

%% file: contents/appendix.tex
\newpage
\input{contents/appendix/app_B_diff_c_rust}
\newpage
\input{contents/appendix/app_C_task_division}

\newpage
\input{contents/appendix/app_D_testing}
\newpage
\input{contents/appendix/app_E_test_harness_example}
\newpage
\input{contents/appendix/app_F_example}
\newpage
\input{contents/appendix/app_G_datasets}
\newpage
\input{contents/appendix/app_H_llm_configs}
\newpage
\input{contents/appendix/app_I_failure_analysis}
\newpage
\input{contents/appendix/app_J_cost}

\newpage
\input{contents/appendix/app_K_ablation}
\newpage
\input{contents/appendix/app_L_temperature}
\newpage
\input{contents/appendix/app_N_spec_rules}
\newpage
\input{contents/appendix/app_O_crust_bench_failures}

\newpage
\input{contents/appendix/app_M_prompts}

%% file: contents/appendix/app_B_diff_c_rust.tex
\section{Differences Between C and Rust}
\label{app:diff_c_rust}

\subsection{Code Snippets}

Here is a code example to demonstrate the differences between C and Rust. The example
shows a simple C program and its equivalent Rust program.
The \texttt{create\_sequence} function takes an integer \texttt{n} as input and
returns an array with a sequence of integers. In C, the function needs to
allocate memory for the array using \texttt{malloc} and will return the pointer
to the allocated memory as an array. If the size is invalid, or the allocation
fails, the function will return \texttt{NULL}. The caller of the function is
responsible for freeing the memory using \texttt{free} when it is done with the
array to prevent memory leaks.

\begin{figure}[h!]
	\scriptsize
	\centering
	\begin{minipage}[t]{0.45\textwidth}
		\textbf{\textsf{C Code:}}
		\roundedinclude[backcolor1]{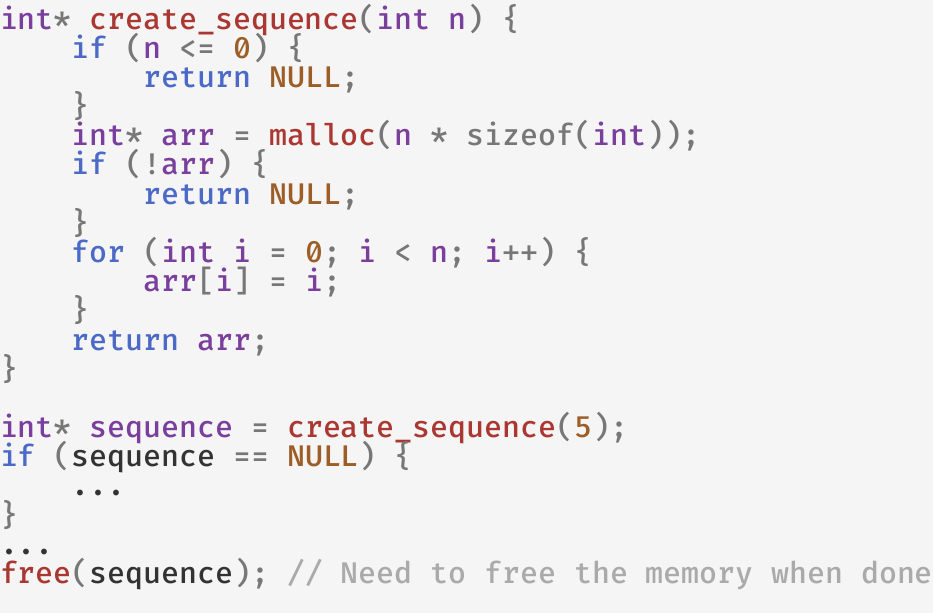}{\textwidth}
    \end{minipage}
    \hfill
    \begin{minipage}[t]{0.45\textwidth}
		\textbf{\textsf{Rust Code:}}
		\roundedinclude[backcolor2]{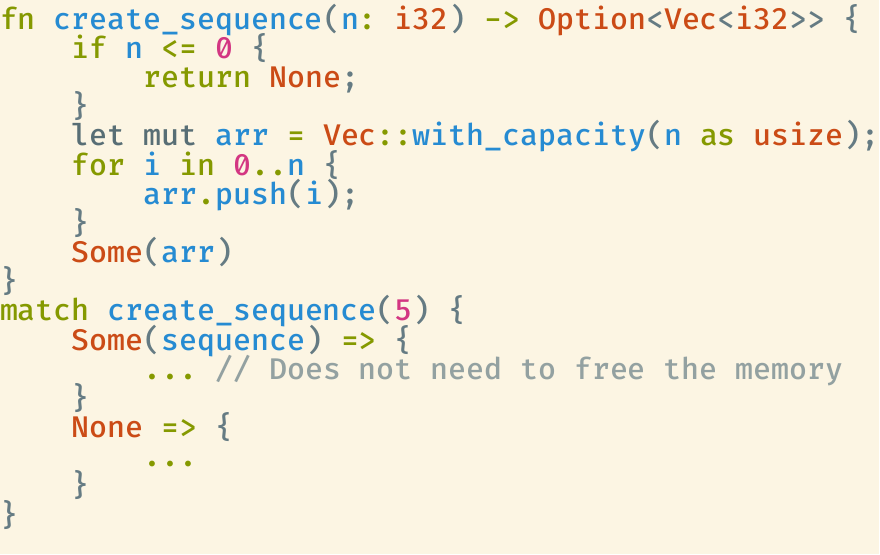}{\textwidth}
	\end{minipage}
	\caption{Example of a simple C program and its equivalent Rust program, both hand-written for illustration.}
	\label{fig:example}
\end{figure}

\subsection{Tabular Summary}

Here, we present a non-exhaustive list of differences between C and Rust in
Table~\ref{tab:diff_c_rust},
highlighting the key features that make translating code from C to Rust
challenging. While the list is not comprehensive, it provides insights into the
fundamental distinctions between the two languages, which can help developers
understand the challenges of migrating C code to Rust.

\begin{table*}[h!]
\centering
\scriptsize

\begin{tabularx}{\textwidth}{>{\scshape}l p{0.3\textwidth} X}
\toprule
\textsc{Feature} & \textsc{C} & \textsc{Rust} \\ \midrule
Memory Management & Manual (through \texttt{malloc/free}) & Automatic (through ownership and borrowing) \\[2ex] %
	Pointers & Raw pointers like \texttt{*p} & Safe references like
	\texttt{\&p/\&mut p},
	\texttt{Box} and \texttt{Rc} \\[2ex] %
		Lifetime Management & Manual freeing of memory & Lifetime annotations and borrow checker\\[2ex] %
Error Handling & Error codes and manual checks & Explicit handling with \texttt{Result} and \texttt{Option} types \\[2ex] %
Null Safety & Null pointers allowed (e.g., \texttt{NULL}) & No null pointers; uses \texttt{Option} for nullable values \\[2ex] %
Concurrency & No built-in protections for data races & Enforces safe concurrency with ownership rules \\[2ex] %
Type Conversion & Implicit conversions allowed and common & Strongly typed; no implicit conversions \\[2ex] %
Standard Library & C stand library with direct system calls & Rust standard library with utilities for strings, collections, and I/O \\[2ex] %
Language Features & Procedure-oriented with minimal abstractions & Modern features like pattern matching, generics, and traits \\[2ex] \bottomrule
\end{tabularx}
\caption{Key Differences Between C and Rust}
\label{tab:diff_c_rust}
\end{table*}

%% file: contents/appendix/app_C_task_division.tex
\section{Preprocessing and Task Division}
\label{app:task_division}
\subsection{Preprocessing of C Files}
To support real-world C projects, \sys parses the compile commands generated by the \texttt{make} tool, extracting relevant flags for preprocessing, parsing, compilation, linking, and third-party tools' use.

C source files usually contain preprocessing directives, such as \texttt{\#include}, \texttt{\#define}, \texttt{\#ifdef}, \texttt{\#endif}, \textit{etc.},  which we need to resolve before parsing C files.
For \texttt{\#include}, we copy and expand non-system headers recursively while keeping \texttt{\#include} of system headers intact, because included non-system  headers contain project-specific definitions such as structs and enums that the LLM has not known while system headers'
 contents are known to the LLM and expanding  them would unnecessarily introduce too much noise. For other directives, we pass relevant C project compile flags to the C preprocessor from GCC to resolve them.

\subsection{Algorithm for Task Division}
The task division algorithm is used to determine the order in which the items
should be translated. The algorithm is shown in Algorithm~\ref{alg:task_division}.

\begin{algorithm*}
	\caption{Translation Task Order Determination}
	\label{alg:task_division}
	\begin{algorithmic}[1]
		\Require $L_i$: List of items to be translated
		\Require $dep(a)$: Function to get dependencies of item $a$
		\Ensure $L_{sorted}$: List of groups resolving dependencies
		\State $L_{sorted} \gets \emptyset$ \Comment{Empty list}
		\While{$|L_{sorted}| < |L_i|$}
		\State $L_{processed} \gets \emptyset$
		\For {$a \in L_{i}$}
		\If{$a \notin L_{processed}$ and $dep(a) \subseteq L_{processed}$}
		\State $L_{sorted} \gets L_{sorted} + a$ \Comment{Add to sorted list}
		\State $L_{processed} \gets L_{processed} \cup a$
		\EndIf
		\EndFor
		\If{$L_{processed} = \emptyset$}
		\State $L_{circular} \gets DFS(L_i, dep)$ \Comment{Circular dependencies}
		\State $L_{sorted} \gets L_{sorted} + L_{circular}$ \Comment{Add a group to sorted list}
		\EndIf
		\EndWhile
		\State \Return $L_{sorted}$
	\end{algorithmic}
\end{algorithm*}

In the algorithm, $L_i$ is the list of items to be translated, and $dep(a)$ is a
function that returns the dependencies of item $a$. The algorithm returns a list
$L_{sorted}$ that contains the items in the order in which they should be
translated. $DFS(L_i, dep)$ is a depth-first search function that returns a list
of items involved in a circular dependency. It begins by collecting all items (e.g., functions, structs) to be
translated and their respective dependencies (in both functions and data types).
Items with no unresolved dependencies are pushed into the translation order list
first, and other items will remove them from their dependencies list. This
process continues until all items are pushed into the list, or circular
dependencies are detected. If circular dependencies are detected, we resolve
them through a depth-first search strategy, ensuring that all items involved in
a circular dependency are grouped together and handled as a single unit.

%% file: contents/appendix/app_D_testing.tex
\section{Equivalence Testing Details in Prior Literature}
\label{app:testing}

\subsection{Symbolic Execution-Based Equivalence}
Symbolic execution explores all potential execution paths of a program by using symbolic inputs to generate constraints~\citep{king1976symbolic, baldoni2018survey, coward1988symbolic}. While theoretically powerful, this method is impractical for verifying C-to-Rust equivalence due to differences in language features. For instance, Rust's RAII (Resource Acquisition Is Initialization) pattern automatically inserts destructors for memory management, while C relies on explicit \texttt{malloc} and \texttt{free} calls. These differences cause mismatches in compiled code, making it difficult for symbolic execution engines to prove equivalence. Additionally, Rust's compiler adds safety checks (e.g., array boundary checks), which further complicate equivalence verification.

\subsection{Fuzz Testing-Based Equivalence}
Fuzz testing generates random or mutated inputs to test whether program outputs match expected results~\citep{zhu2022fuzzing, miller1990empirical, liang2018fuzzing}. While more practical than symbolic execution, fuzz testing faces challenges in constructing meaningful inputs for real-world programs. For example, testing a URL parsing function requires generating valid URLs with specific formats, which is non-trivial. For large C programs, this difficulty scales, making it infeasible to produce high-quality test cases for every translated Rust function.

%% file: contents/appendix/app_E_test_harness_example.tex
\section{An Example of the Test Harness}
\label{app:test_harness_example}

\begin{figure}[h]
	\scriptsize
	\centering
	\begin{minipage}{0.7\linewidth}
		\roundedinclude[backcolor2]{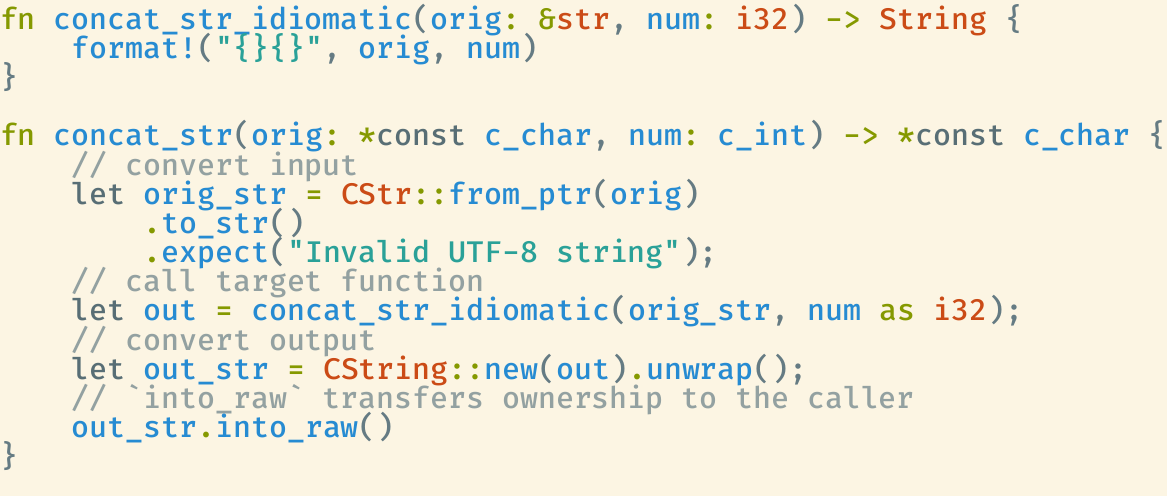}{\textwidth}
	\end{minipage}
	\caption{Test harness used for verifying \texttt{concat\_str} translation}
	\label{fig:harness}
\end{figure}

Here, we provide an example of the test harness used to verify the correctness of the translated code in Figure~\ref{fig:harness}, which is used to verify the idiomatic Rust code.
In this example, the \texttt{concat\_str\_idiomatic} function is the idiomatic translation we are testing, while the \texttt{concat\_str\_c} function is the test harness function that can be linked back to the original C code.
where a string and an integer are passed as input, and an owned string is returned. Input strings are converted from C's \texttt{char*} to Rust's \texttt{\&str}, and output strings are converted from Rust's \texttt{String} back to C's \texttt{char*}.

%% file: contents/appendix/app_F_example.tex
\section{An Example of \sys{} Translation Process}
\label{app:example}

\begin{figure}
	\centering
	\begin{subfigure}{0.55\textwidth}
		\centering
		\roundedinclude[backcolor1]{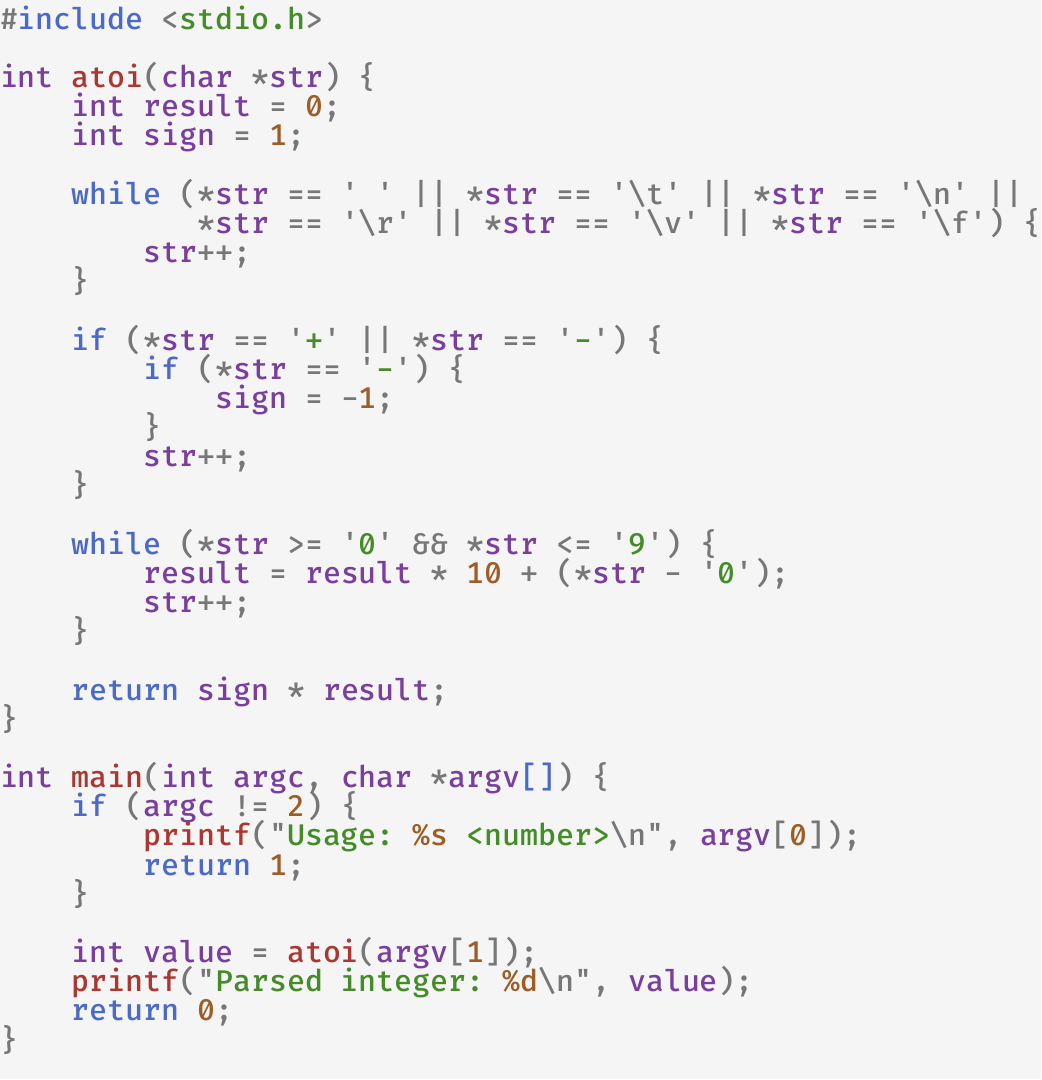}{0.8\textwidth}
		\caption{C implementation of \texttt{atoi}}
		\label{fig:atoi_c}
	\end{subfigure}
	\begin{subfigure}{0.44\textwidth}
		\centering
		\roundedinclude[backcolor3]{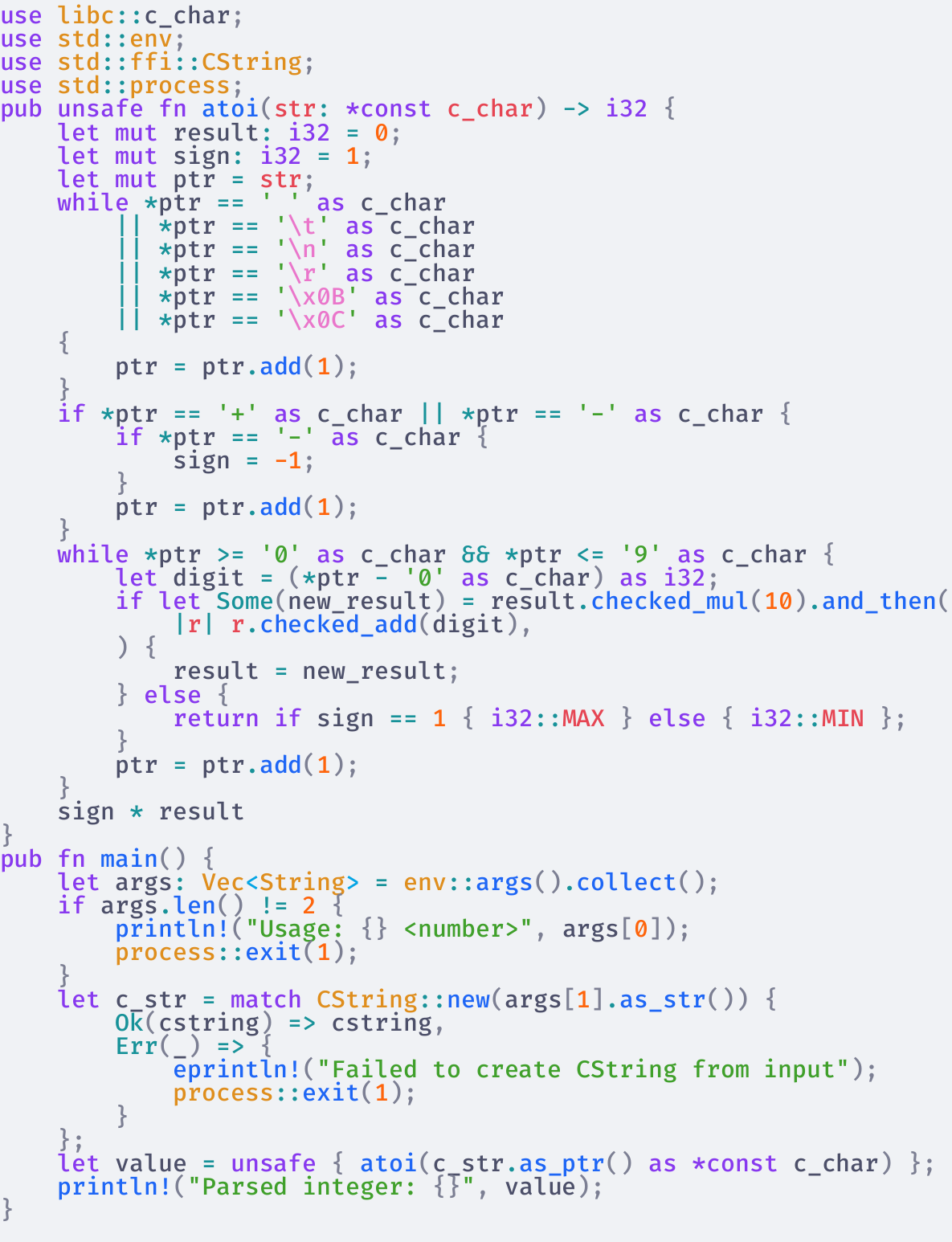}{\textwidth}
		\caption{Unidiomatic Rust translation from C}
		\label{fig:atoi_unidio_rs}
	\end{subfigure}
	\begin{subfigure}{0.6\textwidth}
		\centering
		\roundedinclude[backcolor2]{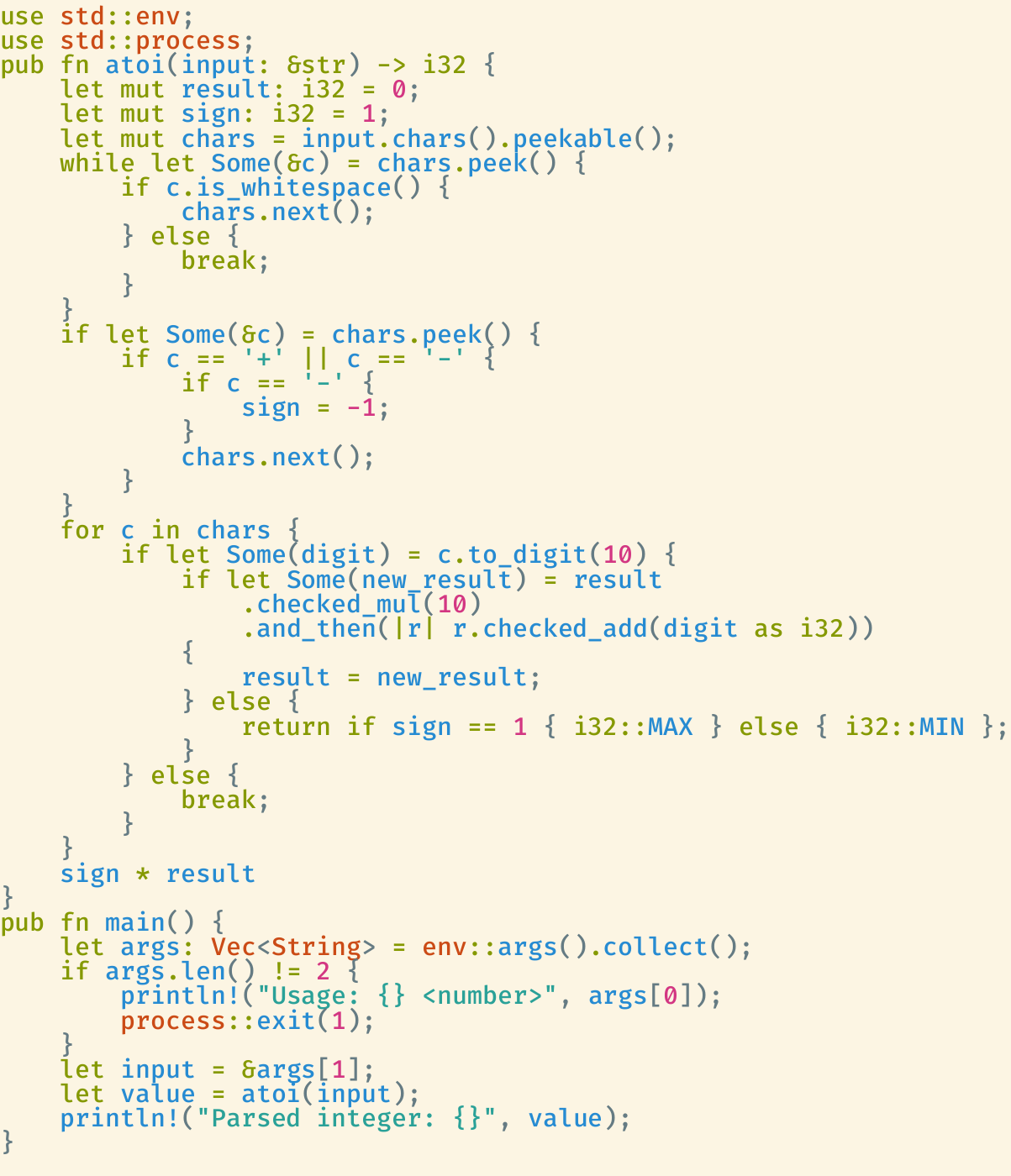}{\textwidth}
		\caption{Idiomatic Rust translation from unidiomatic Rust}
		\label{fig:atoi_idio_rs}
	\end{subfigure}
	\caption{\sys{} translation process for \texttt{atoi} program}
	\label{fig:atoi}
\end{figure}

To demonstrate the translation process of \sys{}, we present a straightforward example of translating a C function to Rust.
The C program includes an \texttt{atoi} function that converts a string to an integer, and a \texttt{main} function that parses command-line arguments and calls the \texttt{atoi} function.
The C code is shown in Figure~\ref{fig:atoi_c}.

We assume that there are numerous end-to-end tests for the C code, allowing
\sys{} to use them for verifying the correctness of the translated Rust code.

First, the divider will divide the C code into two parts: the \texttt{atoi} function and the \texttt{main} function, and determine the translation order is
first \texttt{atoi} and then \texttt{main}, as \texttt{atoi} is the dependency of \texttt{main} and the \texttt{atoi} function is a pure function.

Next, \sys{} proceeds with the unidiomatic translation, converting both
functions into unidiomatic Rust code. This generated code will keep the
semantics of the original C code while using Rust syntax. Once the translation
is complete, the unidiomatic verifier executes the end-to-end tests to ensure
the correctness of the translated function. If the verifier passes all tests,
\sys{} considers the unidiomatic translation accurate and progresses to the next
function. If any test fails, \sys{} will retry the translation process using
the feedback information collected from the verifier, as described in
\S~\ref{sec:verification}. After translating all sections of the C
code, \sys{} will combine the unidiomatic Rust code segments to form the final
unidiomatic Rust code.
The unidiomatic Rust code is shown in Figure~\ref{fig:atoi_unidio_rs}.

Then, the \sys{} will start the idiomatic translation process and translate the
unidiomatic Rust code into idiomatic Rust code. The idiomatic translator
requests the LLM to adapt the C semantics into idiomatic Rust, eliminating any
unsafe and non-idiomatic constructs, as detailed in
\S~\ref{sec:translation}. Based on the same order, the
\sys{} will translate two functions accordingly, and using the idiomatic
verifier to verify and provide the feedback to the LLM if the verification
fails. After all parts of the Rust code are translated into idiomatic Rust,
verified, and combined, the \sys{} will produces the final idiomatic Rust code.
The idiomatic Rust code is shown in Figure~\ref{fig:atoi_idio_rs}, representing the final output of \sys{}.

%% file: contents/appendix/app_G_datasets.tex
\section{Dataset Details}
\label{app:datasets}

\input{figures/dataset_table}

\subsection{TransCoder-IR Dataset~\citep{transcoderir}}

The TransCoder-IR dataset is used to evaluate the TransCoder-IR model and consists of solutions to coding challenges in various programming languages. For evaluation, we focus on the 698 C programs available in this dataset. First, we filter out programs that already have corresponding Rust code. Several C programs in the dataset contain bugs, which are removed by checking their ability to compile. We then use \textit{valgrind} to identify and discard programs with memory errors during the end-to-end tests. Finally, we select 100 programs with the most lines of code for our experiments.

\subsection{Project CodeNet~\citep{codenet}}

Project CodeNet is a large-scale dataset for code understanding and translation,
containing 14 million code samples in over 50 programming languages collected
from online judge websites. From this dataset, which includes more than 750,000
C programs, we target only those that accept external input. Specifically, we
filter programs using \texttt{argc} and \texttt{argv}, which process input from
the command line.
As the end-to-end tests are not available for this dataset,
we develop the \sys{} test generator to automatically generate end-to-end tests
for these programs based on the source code.
For evaluation, we select 200 programs and refine the dataset
to include 100 programs that successfully generate end-to-end tests.

\subsection{CRust-Bench~\citep{khatry2025crust}}

CRust-Bench is a repository-level benchmark for C-to-safe-Rust transpilation. It
collects 100 real-world C repositories (the CBench suite) and pairs each with a
manually written, safe Rust interface and a set of tests that assert functional
correctness. By evaluating full repositories rather than isolated functions,
CRust-Bench surfaces challenges common in practice, such as complex,
pointer-rich APIs. In our evaluation, we use a 50-sample subset in CRust-Bench, which
exclude entries that are out of scope for our pipeline (e.g., circular type or
function dependencies and compiler-specific intrinsics that do not map cleanly).
For each selected sample, we reuse the upstream end-to-end tests and relink
them so that calls exercise our translated code; build environments and link
flags follow the sample's configuration.

\subsection{libogg~\citep{libogg}}

\texttt{libogg} is the reference implementation of the Ogg multimedia container. Ogg is a
stream-oriented format that frames, timestamps, and multiplexes compressed media
bitstreams (e.g., audio/video) into a robust, seekable stream. The \texttt{libogg}
distribution contains only the Ogg container library (codecs such as Vorbis or
Theora are hosted separately). In our case study, the codebase comprises roughly
2{,}041 lines of code (excluding tests), six struct definitions, three global
variables, and 77 exported functions. We use the project's upstream tests and
build scripts. This single-project evaluation complements the CRust-Bench
subset by focusing on non-trivial structs, buffers, and pointer manipulation in
a real-world C library.

%% file: figures/dataset_table.tex
\begin{table*}[h]
    \centering
    \scriptsize
    \label{tab:datasets}
    \begin{tabularx}{\textwidth}{>{\scshape}l r X c c c}
        \toprule
        \textsc{Dataset} & \textsc{Size} & \textsc{Preprocessing} & \textsc{E2E Tests} & \textsc{Coverage (Line/Func)} \\ \midrule
        \textsc{TransCoder-IR}~\citep{transcoderir} & 100 & Removed buggy programs (compilation/memory errors) and entries with existing Rust & Present & 97.97\% / 99.5\% \\
        \textsc{Project CodeNet}~\citep{codenet} & 100 & Filtered for external-input programs (\texttt{argc}/\texttt{argv}); auto-generated tests & Generated & 94.37\% / 100\% \\
        \textsc{CRust-Bench}~\citep{khatry2025crust} & 50 & Excluded unsupported patterns; combine code of each sample to a single \texttt{lib.c} & Present & 76.18\% / 80.98\% \\
        \textsc{libogg}~\citep{libogg} & 1 & None. Each component of the library is contained within a single C file. & Present & 83.3\% / 75.3\% \\
        \bottomrule
    \end{tabularx}
    \caption{Summary of datasets and real-world code-bases used for evaluation; coverage audited with \texttt{gcov} on the tests exercised in our pipeline.}
\end{table*}

%% file: contents/appendix/app_H_llm_configs.tex
\section{LLM Configurations}
\label{app:llm_config}

Table~\ref{tab:llm_configurations} shows our configurations for different LLMs in evaluation. All other hyperparameters (e.g., Top-P, Top-K) use provider defaults.
As GPT-5 does not support temperature setting, we use its default temperature.

\begin{table}[h]
	\centering
	\scriptsize
	\begin{threeparttable}
		\begin{tabular}{lll}
			\toprule
			\textbf{Model}         & \textbf{Version}                & \textbf{Temperature} \\
			\midrule
			GPT-4o                 & gpt-4o-2024-08-06               & 0                    \\
			Claude 3.5 Sonnet      & claude-3-5-sonnet-20241022      & 0                    \\
			Gemini 2.0 Flash       & gemini-2.0-flash-exp            & 0                    \\
			Llama 3.3 Instruct 70B & Llama 3.3 Instruct 70B\tnote{1} & 0                    \\
			DeepSeek-R1            & DeepSeek-R1 671B\tnote{2}       & 0                    \\
			GPT-5                  & gpt-5-2025-08-07                & default              \\
			\bottomrule
		\end{tabular}
		\begin{tablenotes}
			\item[1] https://huggingface.co/meta-llama/Llama-3.3-70B-Instruct
			\item[2] https://huggingface.co/deepseek-ai/DeepSeek-R1
		\end{tablenotes}
	\end{threeparttable}
	\caption{Configurations of Different LLMs in Evaluation}
	\label{tab:llm_configurations}
\end{table}

%% file: contents/appendix/app_I_failure_analysis.tex
\section{Failure Analysis in Evaluating \sys{}}
\label{subsub:failure_analysis}

\begin{table*}[ht]
    \scriptsize
    \centering
    \begin{subtable}[t]{\textwidth}
        \caption{TransCoder-IR}
        \label{tab:failure_reasons_transcoder}
        \centering
        \begin{tabularx}{\textwidth}{>{\centering\arraybackslash\scshape}p{1.3cm}>{\arraybackslash}X}
            \toprule
            \textsc{Category} & \textsc{Description} \\
            \midrule
            R1 & Memory safety violations in array operations due to improper bounds checking \\
            R2 & Mismatched data type translations \\
            R3 & Incorrect array sizing and memory layout translations \\
            R4 & Incorrect string representation conversion between C and Rust \\
            R5 & Failure to handle C's undefined behavior with Rust's safety mechanisms \\
            R6 & Use of C-specific functions in Rust without proper Rust wrappers \\
            \bottomrule
        \end{tabularx}
    \end{subtable}
    \begin{subtable}[t]{\textwidth}
        \caption{Project CodeNet}
        \label{tab:failure_reasons_codenet}
        \centering
        \begin{tabularx}{\textwidth}{>{\centering\arraybackslash\scshape}p{1.3cm}>{\arraybackslash}X}
            \toprule
            \textsc{Category} & \textsc{Description} \\
            \midrule
            S1 & Improper translation of command-line argument handling or attempt to fix wrong handling \\
            S2 & Function naming mismatches between C and Rust \\
            S3 & Format string directive mistranslation causing output inconsistencies \\
            S4 & Original code contains random number generation \\
            S5 & \sys{} unable to translate mutable global state variables \\
            S6 & Mismatched data type translations \\
            S7 & Incorrect control flow or loop boundary condition translations \\
            \bottomrule
        \end{tabularx}
    \end{subtable}
    \caption{Failure reason categories for translating TransCoder-IR and Project CodeNet datasets.}
    \label{tab:failure_reasons}
\end{table*}

\begin{figure*}[ht]
    \vspace{-4mm}
    \centering
    \includegraphics[width=0.8\linewidth]{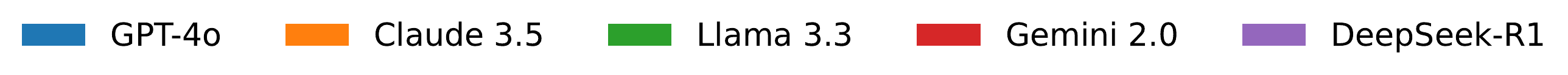}
    \begin{subfigure}[t]{0.52\textwidth}
        \centering
        \includegraphics[width=\linewidth]{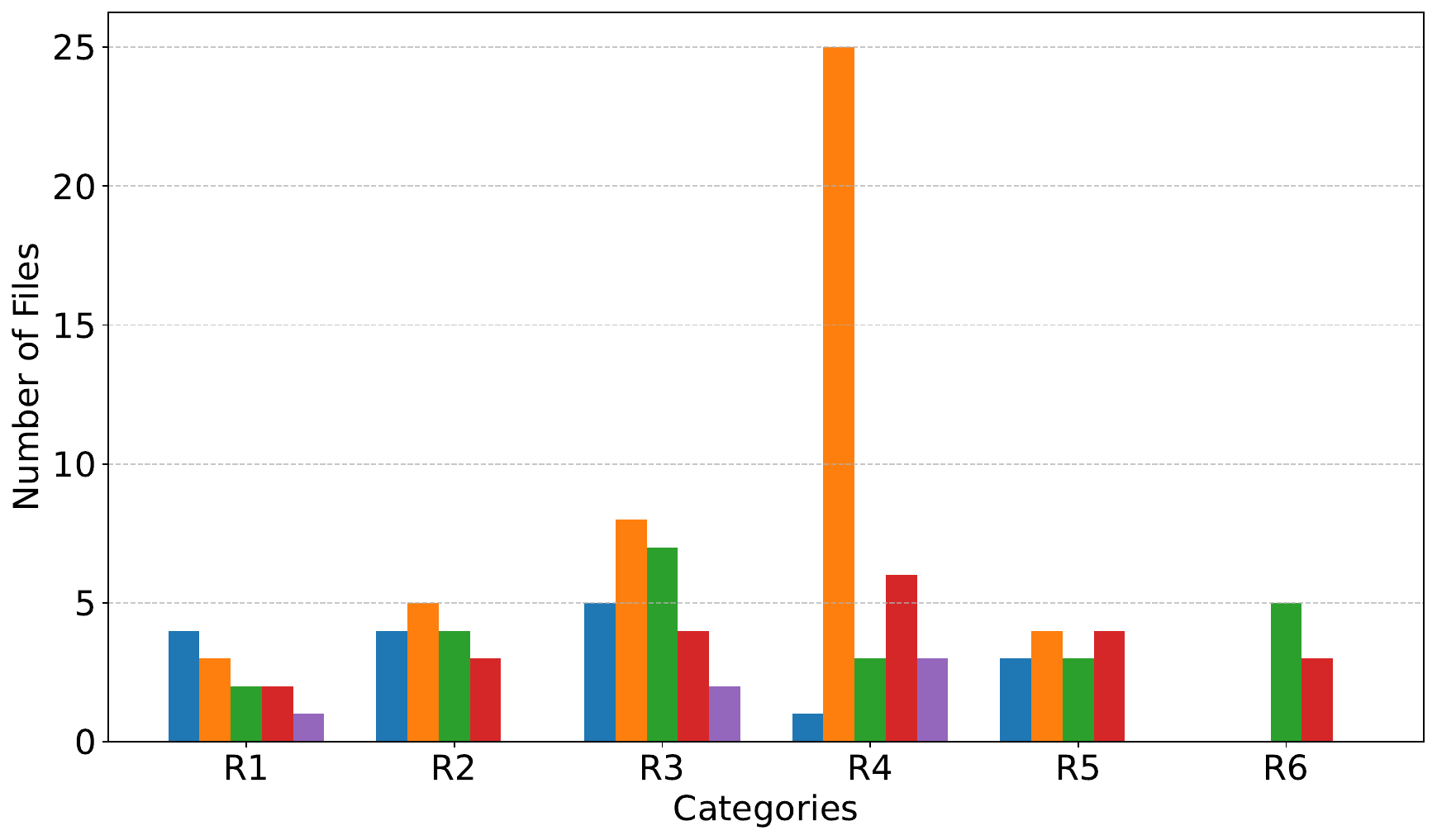}
        \caption{TransCoder-IR}
        \label{fig:failure_reasons_transcoder}
    \end{subfigure}
    \hfill
    \begin{subfigure}[t]{0.47\textwidth}
        \centering
        \includegraphics[width=\linewidth]{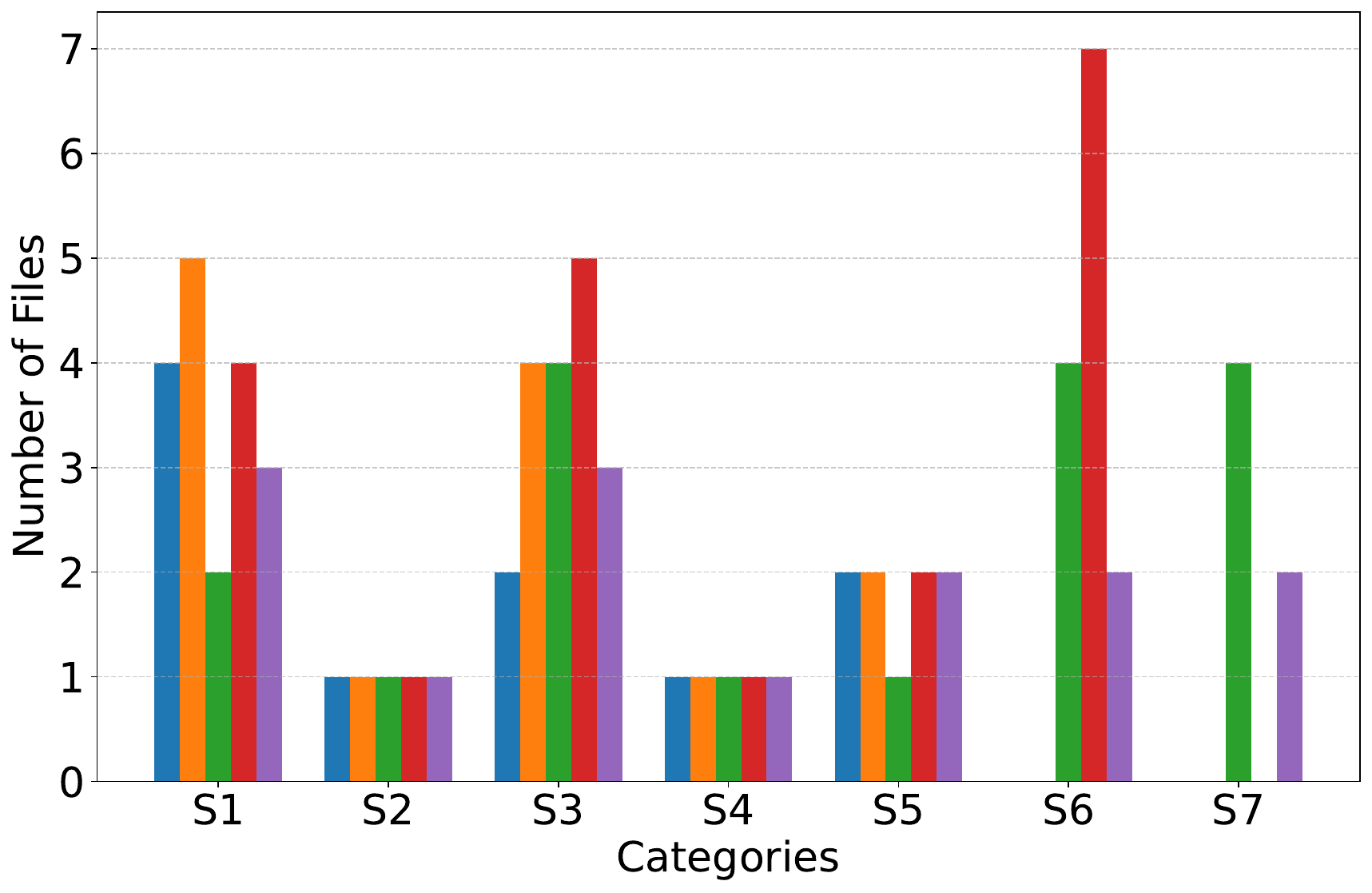}
        \caption{Project CodeNet}
        \label{fig:failure_reasons_codenet}
    \end{subfigure}
    \caption{Failure reasons across different LLM models for both datasets.}
    \label{fig:failure_reasons_combined}
\end{figure*}

Here, we analyze the failure cases of \sys{} in translating C code to Rust that
we conducted in Section~\ref{subsub:success_rate}.
as cases where \sys{} fails offer valuable insights into areas that require
refinement. For each failure case in the two datasets, we conduct an analysis to
determine the primary cause of translation failure. This process involves leveraging DeepSeek-R1 to identify potential reasons (prompts available in Appendix~\ref{app:failure-reason-analysis}), followed by
manual verification to ensure correctness.
We only focus on the translation process
from C to unidiomatic Rust because: (1) it is the most challenging step, and (2)
it can better reflect the model's ability to fit the syntactic and semantic
differences between the two languages.
Table~\ref{tab:failure_reasons}
summarize the categories of failure reasons, and
Figure~\ref{fig:failure_reasons_transcoder}
and~\ref{fig:failure_reasons_codenet}
illustrate failure reasons (FRs) across models.

\noindent{\bf (1) TransCoder-IR} (Table~\ref{tab:failure_reasons_transcoder},
Figure~\ref{fig:failure_reasons_transcoder}):
Based on the
analysis, we observe that different models exhibit varying failure
reasons. Claude 3.5 shows a particularly high incidence of string representation
conversion errors (R4), with 25 out of 45 total failures in the unidiomatic
translation step. In contrast, GPT-4o has only 1 out of 17 failures in this
category. Llama 3.3 demonstrates consistent challenges with both R3 (incorrect array sizing and memory layout
translations) and R6 (using C-specific functions without proper Rust wrappers),
with 10 files for each category. GPT-4o shows a more balanced distribution of
errors, with its highest count in R3. All models except GPT-4o struggle with string handling (R4) to
varying degrees, suggesting this is one of the most challenging aspects of the
translation process. For R6 (use of C-specific functions in Rust), which primarily is a compilation failure,
only Llama 3.3 and Gemini 2.0 consistently fail to resolve the issue in some cases, while all other models can successfully handle the compilation errors through feedback
and avoid failure in this category.
DeepSeek-R1 has the fewest overall errors across categories,
with failures only in R1 (1 file), R3 (2 files), and R4 (3 files), while completely
avoiding errors in R2, R5, and R6.

\noindent{\bf (2) Project CodeNet} (Table~\ref{tab:failure_reasons_codenet},
Figure~\ref{fig:failure_reasons_codenet}):
Similar to the TransCoder-IR dataset, we also observe that different models in
Project CodeNet demonstrate varying failure reasons. C-to-Rust code translation
challenges in the CodeNet dataset. Most notably, S6 (mismatched data type
translations) presents a significant barrier for Llama 3.3 and Gemini 2.0 (7
files each), while GPT-4o and Claude 3.5 completely avoid this issue. Input
argument handling (S1) and format string mistranslations (S3) emerge as common
challenges across all models in CodeNet, suggesting fundamental difficulties in
translating these language features regardless of model architecture. Only
Llama 3.3 and DeepSeek-R1 encounter control flow translation failures (S7),
with 2 files each. S4 (random number generation) and S5 (mutable global state
variables) are unable to be translated by \sys{} because the current \sys{}
implementation does not support these features.

Compared to the results in TransCoder-IR, string representation conversion (R4 in
TransCoder-IR, S3 in CodeNet)  remains a consistent challenge across both
datasets for all models, though the issue is significantly more severe in
TransCoder-IR, particularly for Claude 3.5 (24 files). This also suggests that
reasoning models like DeepSeek-R1 are better at handling complex code
logic and string/array manipulation, as they exhibit fewer failures in these areas,
demonstrating the potential of reasoning models to address complex translation tasks.

%% file: contents/appendix/app_J_cost.tex
\section{\sys{} Cost Analysis}
\label{subsec:cost}

\input{figures/cost_table}

Here, we conduct a cost analysis of \sys{} for experiments in \S~\ref{subsub:success_rate} to
evaluate the efficiency of different LLMs in generating idiomatic Rust code.
To evaluate the cost of our approach, we measure (1) {\em Total LLM Queries} as the number of total LLM queries made during translation and verification for a single test case in each dataset, and %
(2) {\em Total Token Count} as the total number of tokens processed by the LLM for a single test case in each dataset.
To ensure a fair comparison across models, we use the same tokenizer (\texttt{tiktoken}) and encoding (\texttt{o200k\_base}).

In order to better understand costs, we only analyze programs that successfully generate idiomatic Rust code, excluding failed attempts (as they always reach the maximum retry limit and do not contribute meaningfully to the cost analysis). We evaluate the combined cost of both translation phases to assess overall efficiency. Table~\ref{tab:llm_cost_analysis} compares the average cost of different LLMs across two datasets, measured in token usage and query count per successful idiomatic Rust translation as mentioned in \S~\ref{subsec:metrics}.

\noindent{\bf Results:} Gemini 2.0 and GPT-4o are the most efficient models,
requiring the fewest tokens and queries. GPT-4o maintains a low token cost
(2651.21 on TransCoder-IR, 2565.36 on CodeNet) with 4.24 and 2.95 average
queries, respectively. Gemini 2.0 is similarly efficient, especially on CodeNet,
with the lowest token usage (2209.38) and requiring only 2.39 queries on
average. Claude 3.5, despite its strong performance on CodeNet, incurs higher
costs on TransCoder-IR (4595.33 tokens, 5.15 queries), likely due to additional
translation steps. Llama 3.3 is the least efficient in non-thinking model (GPT-4o,
Claude 3.5, Gemini 2.0), consuming the most tokens (4622.80 and 4456.84, respectively) and requiring the highest number of queries (5.39 and 3.80, respectively), indicating significant resource demands.

As a reasoning model, DeepSeek-R1 consumes significantly more
tokens (17,895.52 vs. 13,592.61) than non-reasoning models--5-7 times
higher than GPT-4o--despite having a
similar average query count (4.77 vs. 3.11) for generating idiomatic Rust code. This high token usage comes from the ``reasoning process'' required before code
generation.

%% file: figures/cost_table.tex
\begin{table}[h]
\centering
\scriptsize
\begin{tabular}{l c c c c c}
\toprule
\textsc{LLM} & \textsc{Dataset}  & \textsc{Tokens} &  \textsc{Avg. Queries} \\
\midrule
Claude 3.5 & TransCoder-IR & \cellcolor{RedOrange!30} 4595.33 & \cellcolor{RedOrange!83} 5.15 \\
 & CodeNet & \cellcolor{RedOrange!26} 3080.28 & \cellcolor{RedOrange!63} 3.15 \\
\midrule
Gemini 2.0 & TransCoder-IR & \cellcolor{RedOrange!24} 3343.12 & \cellcolor{RedOrange!20} 4.24 \\
 & CodeNet & \cellcolor{RedOrange!20} 2209.38 & \cellcolor{RedOrange!20} 2.39 \\
\midrule
Llama 3.3 & TransCoder-IR & \cellcolor{RedOrange!30} 4622.80 & \cellcolor{RedOrange!100} 5.39 \\
 & CodeNet & \cellcolor{RedOrange!36} 4456.84 & \cellcolor{RedOrange!100} 3.80 \\
\midrule
GPT-4o & TransCoder-IR & \cellcolor{RedOrange!20} 2651.21 & \cellcolor{RedOrange!20} 4.24 \\
 & CodeNet & \cellcolor{RedOrange!23} 2565.36 & \cellcolor{RedOrange!52} 2.95 \\
\midrule
DeepSeek-R1 & TransCoder-IR & \cellcolor{RedOrange!100} 17895.52 & \cellcolor{RedOrange!57} 4.77 \\
 & CodeNet & \cellcolor{RedOrange!100} 13592.61 & \cellcolor{RedOrange!61} 3.11 \\
\bottomrule
\end{tabular}
\caption{Average Cost Comparison of Different LLMs Across Two Datasets. The color intensity represents the relative cost of each metric for each dataset.}
\label{tab:llm_cost_analysis}
\end{table}

%% file: contents/appendix/app_K_ablation.tex
\section{Ablation Study on \sys{} Designs}
\label{app:ablation}

\noindent This appendix reports additional ablations that evaluate key design
choices in \sys{}.
All experiments in this section use GPT-4o with the same configuration as
Table~\ref{tab:llm_configurations}.

\subsection{Feedback Mechanism}
\label{app:ablation_feedback}

To evaluate the effectiveness of the feedback mechanism proposed in \S~\ref{subsec:feedback}, we conduct an ablation study by removing the mechanism and comparing the model's performance with and without it. We consider two experimental groups: (1) with the feedback mechanism enabled, and (2) without the feedback mechanism. In the latter setting, if any part of the translation fails, the system simply restarts the translation attempt using the original prompt, without providing any feedback from the failure.

We use the same dataset and evaluation metrics described in \S~\ref{sec:setup}, and focus our evaluation on only two models: GPT-4o and Llama 3.3 70B. We choose these models because GPT-4o demonstrated one of the highest performance and Llama 3.3 70B the lowest in our earlier experiments. By comparing the success rates between the two groups, we assess whether the feedback mechanism improves translation performance across models of different capabilities.

\begin{figure*}[ht]
    \centering
    \includegraphics[width=0.95\linewidth]{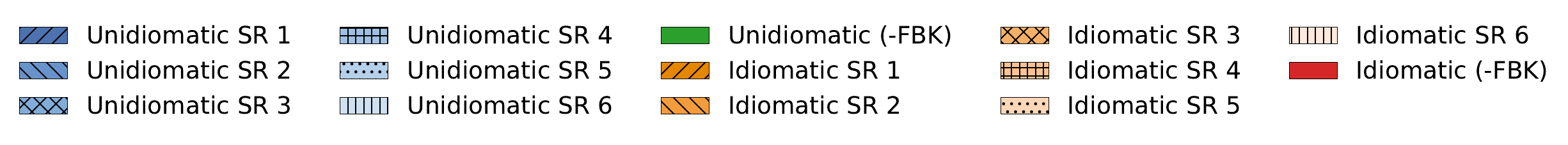}
    \begin{subfigure}{0.49\linewidth}
        \centering
        \includegraphics[width=\linewidth]{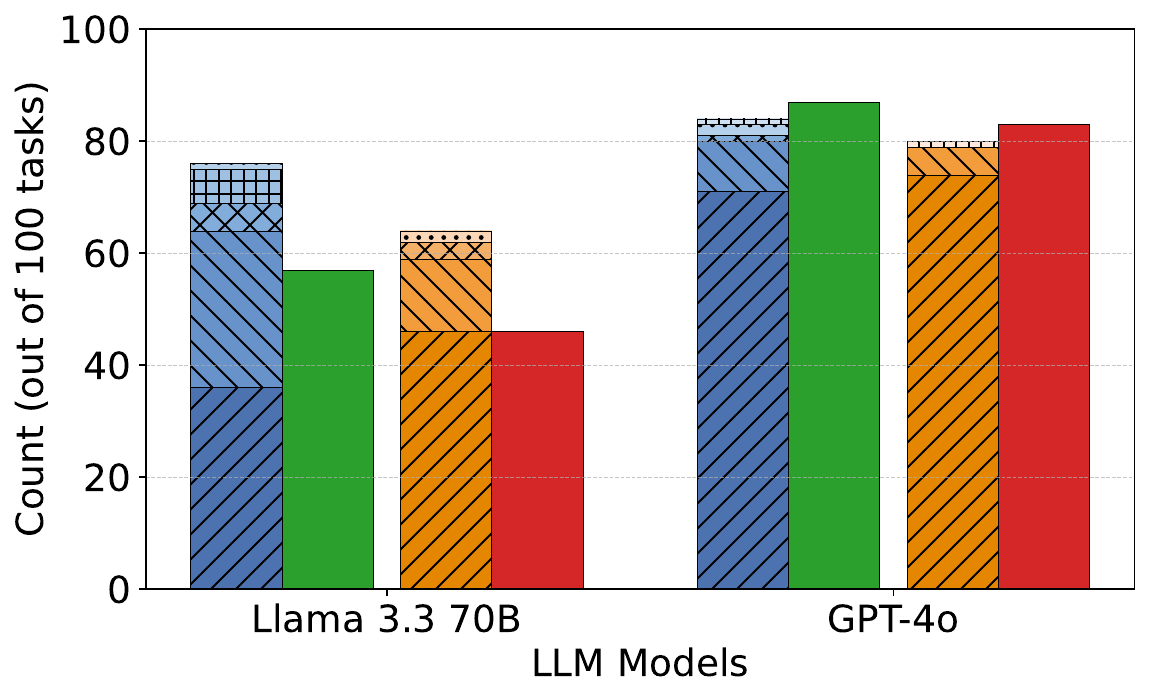}
        \caption{TransCoder-IR With/Without Feedback}
        \label{fig:transcoder_feedback}
    \end{subfigure}
    \hfill
    \begin{subfigure}{0.49\linewidth}
        \centering
        \includegraphics[width=\linewidth]{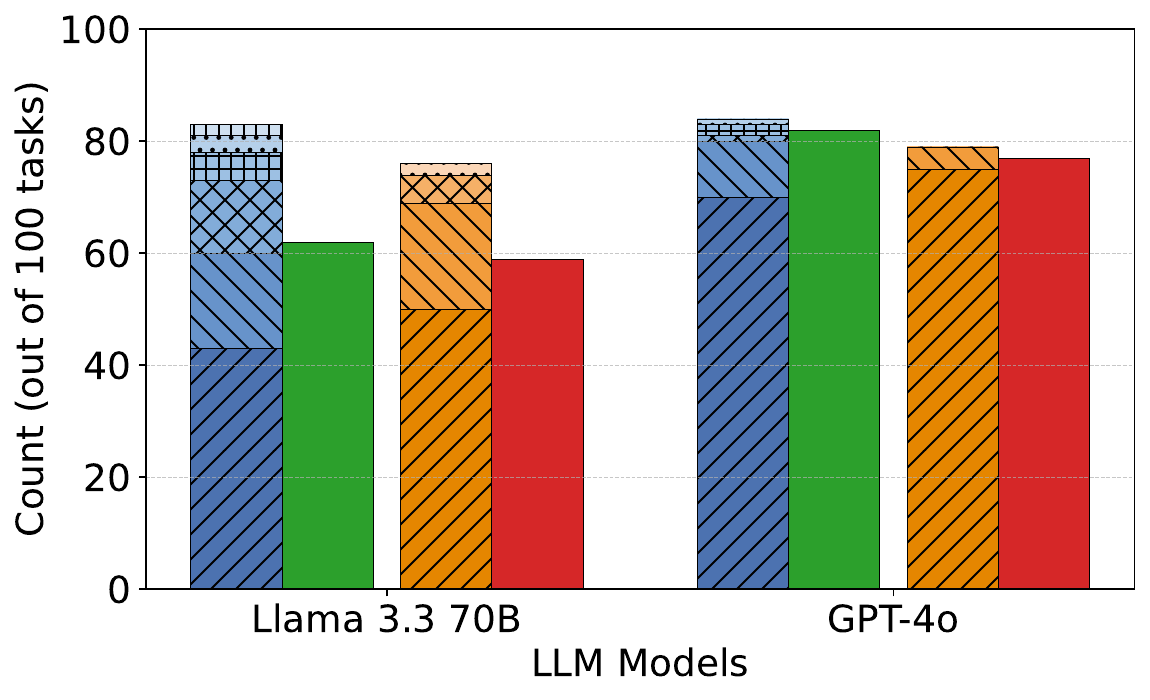}
        \caption{CodeNet With/Without Feedback}
        \label{fig:codenet_feedback}
    \end{subfigure}
    \caption{Ablation study on the feedback mechanism. The success rates of the models with and without the feedback (marked as \textit{-FBK})
    mechanism are shown for both TransCoder-IR and CodeNet datasets.}
    \label{fig:feedback}
\end{figure*}

The results are shown in Figure~\ref{fig:feedback}.

\textbf{(1) TransCoder-IR} (Figure~\ref{fig:transcoder_feedback}):
Incorporating the feedback mechanism increased the number of successful
translations for Llama 3.3 70B from 57 to 76 in the unidiomatic setting and
from 46 to 64 in the idiomatic setting. In contrast, GPT-4o performed slightly
worse with feedback, decreasing from 87 to 84 (unidiomatic) and from 83 to 80
(idiomatic).

\textbf{(2) Project CodeNet} (Figure~\ref{fig:codenet_feedback}): A similar trend
is observed where Llama 3.3 70B improved from 62 to 83 (unidiomatic) and from
59 to 76 (idiomatic), corresponding to gains of 21 and 17 percentage points,
respectively. GPT-4o, however, showed only marginal improvements: from 82 to 84
in the unidiomatic setting and from 77 to 79 in the idiomatic setting.

These results suggest that the feedback mechanism is particularly effective for
lower-capability models like Llama 3.3, substantially improving their
translation success rates. In contrast, higher-capability models such as GPT-4o
already perform near optimal with simple random sampling, leaving little space for
improvement. This indicates that the feedback mechanism is more beneficial for
models with lower capabilities, as they can leverage the feedback to enhance their
overall performance.

\subsection{Plain LLM Translation vs. \sys{}}
\label{app:ablation_plain_llm}

We compare \sys{} against a trivial baseline where GPT-4o directly translates each CRust-Bench sample from C to Rust in a single step. We reuse the same end-to-end (E2E) test harness as \sys{}, and give the trivial baseline more budget: up to 10 repair attempts with compiler/test feedback (vs. 6 attempts in \sys{}).
We study two prompts: (i) a minimal one (``translate the following C code to Rust''); and (ii) an interface-preserving one that explicitly asks the model to preserve pointer arithmetic, memory layout, and integer type semantics (thereby encouraging \texttt{unsafe}).
We report \emph{function success} as the fraction of functions whose Rust translation passes all tests, and \emph{sample success} as the fraction of samples where all translated functions pass.

\begin{table*}[h]
\centering
\scriptsize
\begin{tabular}{lccccc}
\toprule
\textsc{Method} & \textsc{Max att.} & \textsc{Function success} & \textsc{Sample success} & \textsc{Avg. Clippy / func.} \\
\midrule
\sys{} unidiomatic & 6 & \textbf{788/966 (81.57\%)} & 32/50 (64.00\%) & 2.96 \\
\sys{} idiomatic$^\dagger$ & 6 & \textbf{249/580 (42.93\%)} & 8/32 (25.00\%) & 0.28 \\
\midrule
Trivial (1-step) & 10 & 77/966 (7.97\%) & 12/50 (24.00\%) & 1.60 \\
Trivial (1-step, encourage \texttt{unsafe}) & 10 & 207/966 (21.43\%) & 20/50 (40.00\%) & 1.90 \\
\bottomrule
\end{tabular}
\caption{Plain LLM translation vs. \sys{} on CRust-Bench (GPT-4o). The trivial baselines directly translate each sample in one step with up to 10 repair attempts. $\dagger$ The idiomatic stage is evaluated only on samples whose unidiomatic stage fully translated all functions.}
\label{tab:ablation_plain_llm}
\end{table*}

\noindent\textbf{Results on \texttt{CRust-Bench}.} Even with 10 attempts and an ``encourage \texttt{unsafe}'' prompt, the trivial baseline reaches only 21.43\% function success and 40.00\% sample success. Its sample-level performance exceeds \sys{}'s idiomatic stage (40.00\% vs. 25.00\%) because preserving C-style pointer logic in \texttt{unsafe} Rust is substantially easier than performing an idiomatic rewrite. However, \sys{} achieves much higher function-level correctness and produces significantly more idiomatic code (e.g., 0.28 vs. 1.90 average Clippy alerts per function).

\noindent\textbf{Results on \texttt{libogg}.} Under the same E2E tests and attempt budget as \sys{}, both trivial prompts fail to produce any test-passing translations, whereas \sys{} achieves 100\% unidiomatic and 53\% idiomatic success with GPT-4o (Table~\ref{tab:libogg-eval}). This indicates that plain one-shot translation collapses on pointer-heavy libraries, while \sys{} remains effective.

\subsection{Effect of Crown in the Idiomatic Stage}
\label{app:ablation_crown}

We ablate Crown's contribution to idiomatic translation (\S~\ref{sec:translation}) on \texttt{libogg}, using the same setup as \S~\ref{subsec:complex} and keeping all other components unchanged. Table~\ref{tab:ablation_crown} reports idiomatic function success with and without Crown.

\begin{table}[h]
\centering
\scriptsize
\begin{tabular}{lccc}
\toprule
\textsc{Configuration} & \textsc{\# idiom.-successful funcs.} & \textsc{Idiom. SR} & \textsc{Rel. drop} \\
\midrule
\sys{} & 41 & 53\% & -- \\
\sys{} w/o Crown & 34 & 44\% & \textbf{17\%} \\
\bottomrule
\end{tabular}
\caption{Ablating Crown on \texttt{libogg} (GPT-4o).}
\label{tab:ablation_crown}
\end{table}

\noindent\textbf{Results and Representative failure patterns.} Turning off Crown reduces idiomatic success from 41 to 34 functions. The failures are structured. Two representative patterns are:
\begin{lstlisting}[language=Rust]
// Without Crown (shape lost):
pub struct OggPackBuffer { pub ptr: usize }

// With Crown (shape preserved):
pub struct OggPackBuffer { pub ptr: Vec<u8> }

// Without Crown (ownership misclassified as owned):
pub struct OggIovec { pub iov_base: Vec<u8> }

// With Crown (ownership made explicit):
pub struct OggIovec<'a> { pub iov_base: &'a [u8] }
\end{lstlisting}

Once a buffer pointer is collapsed into a scalar index, the harness cannot reconstruct a valid C-facing view of the struct, so pointer arithmetic and buffer access fail together. Similarly, if a non-owning pointer (e.g., \texttt{unsigned char *iov\_base}) is misclassified as owned storage (\texttt{Vec<u8>}), Rust ends up ``owning'' memory that C actually controls, making safe round-tripping infeasible without inventing allocation/free rules that do not exist.

\noindent\textbf{Interpretation.} These failures do not indicate model weakness but an \emph{information-theoretic limitation}: local C syntax does not encode pointer fatness or ownership. For a declaration such as \texttt{char *iov\_base}, both \texttt{Vec<u8>} and \texttt{\&mut u8} are locally plausible. Even an \textit{idealized oracle model} cannot uniquely infer the correct Rust type without global information about ownership and fatness. Crown supplies these semantics via whole-program static analysis; removing it makes idiomatic translation of pointer-heavy code underdetermined and explains the observed drop.

\subsection{Prompting about \texttt{unsafe} in Stage 1}
\label{app:ablation_unsafe_prompt}

We ablate the stage-1 (unidiomatic translation) prompt line that says ``the model may use \texttt{unsafe} if needed.'' All experiments in this subsection are conducted on \texttt{libogg}, using exactly the same setup as in \S~\ref{subsec:complex}.

\subsubsection{Removing ``may use \texttt{unsafe} if needed''}

\noindent We compare the original stage-1 prompt with a variant that deletes this line, keeping everything else unchanged.

\begin{table*}[h]
\centering
\scriptsize
\begin{tabular}{lccccc}
\toprule
\textsc{Prompt variant} & \textsc{Unidi. SR} & \textsc{Clippy total} & \texttt{missing\_safety\_doc} & \texttt{not\_unsafe\_ptr\_arg\_deref} & \textsc{Unsafe fraction} \\
\midrule
Baseline stage 1 (may use \texttt{unsafe}) & 100\% & 108 & 76 & 1 & 8704/8705 (99.99\%) \\
Remove ``may use \texttt{unsafe}'' & 100\% & \textbf{224} & 37 & \textbf{146} & 8100/8219 (98.55\%) \\
\bottomrule
\end{tabular}
\caption{Removing explicit permission to use \texttt{unsafe} in stage 1 on \texttt{libogg} (GPT-4o).}
\label{tab:ablation_unsafe_a3a}
\end{table*}

\noindent Two observations follow. \emph{(1) Overall unsafety hardly changes:} the unsafe fraction drops only from 99.99\% to 98.55\%. \emph{(2) The safety profile becomes worse:} \texttt{clippy::not\_unsafe\_ptr\_arg\_deref} jumps from 1 to 146. That is, the model keeps APIs safe-looking but dereferences raw pointer arguments inside function bodies, pushing unsafety from explicit \texttt{unsafe fn} signatures into hidden dereferences inside safe-looking public functions.

\subsubsection{Replacing With ``AVOID using \texttt{unsafe}''}

\noindent We replace ``may use \texttt{unsafe} if needed'' with a stronger directive: ``AVOID using \texttt{unsafe} whenever possible''.

\begin{table}[h]
\centering
\scriptsize
\begin{tabular}{lcccc}
\toprule
\textsc{Prompt variant} & \textsc{Passed/Total} & \textsc{SR} & \textsc{Rel. drop} \\
\midrule
Baseline stage 1 & 77/77 & 100\% & -- \\
Replace with ``AVOID \texttt{unsafe}'' & 66/77 & 85\% & \textbf{15\%} \\
\bottomrule
\end{tabular}
\caption{Discouraging \texttt{unsafe} in stage 1 harms unidiomatic success on \texttt{libogg} (GPT-4o).}
\label{tab:ablation_unsafe_a3b}
\end{table}

\noindent Under ``AVOID \texttt{unsafe}'', the model often attempts premature ``safe Rust'' rewrites of pointer-heavy C code (changing buffer layouts, index arithmetic, and integer types), which increases logic and type errors and breaks translations. Together, these two prompt variants show that discouraging \texttt{unsafe} in stage 1 harms correctness and produces a worse safety profile, supporting our design choice: allow necessary \texttt{unsafe} in the syntactic first stage, then systematically remove it in the idiomatic refinement stage.

%% file: contents/appendix/app_L_temperature.tex
\section{\sys{} Performance with Different Temperatures}
\label{app:temperature}

\begin{figure*}[h]
	\centering
	\includegraphics[width=0.8\textwidth]{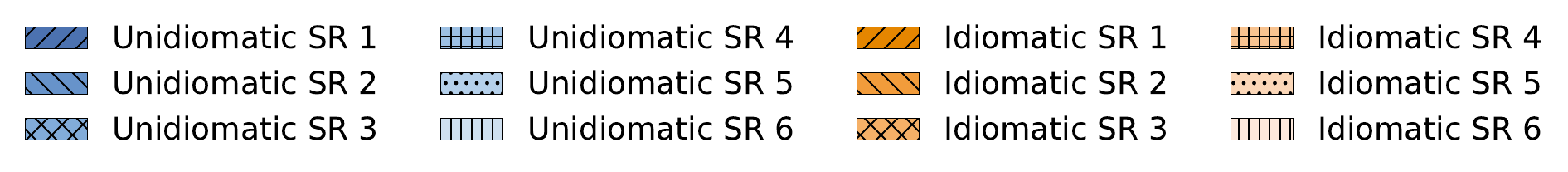}
	\begin{subfigure}{0.49\textwidth}
		\centering
		\includegraphics[width=\textwidth]{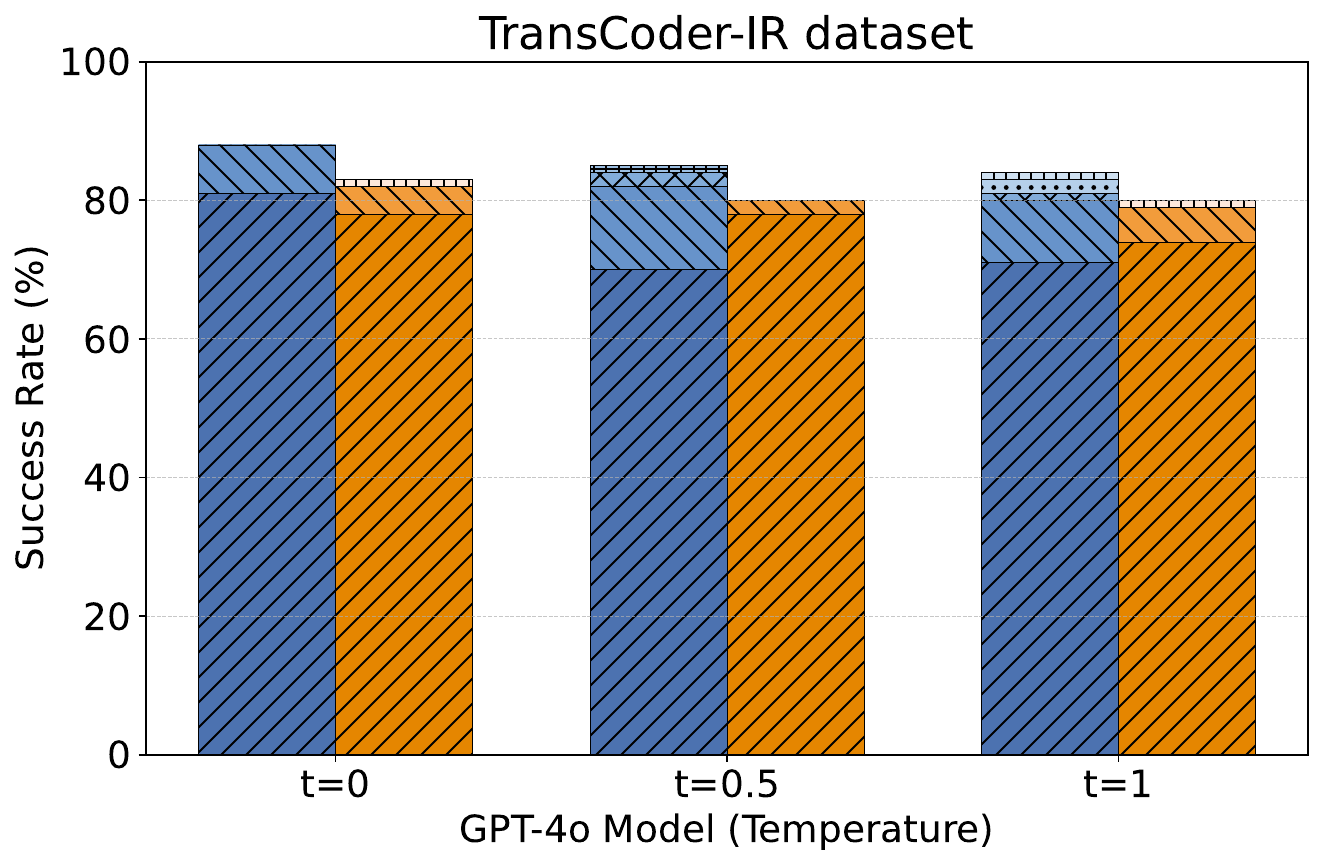}
		\caption{Success Rate on TransCoder-IR}
		\label{fig:temp_transcoder}
	\end{subfigure}
	\begin{subfigure}{0.49\textwidth}
		\centering
		\includegraphics[width=\textwidth]{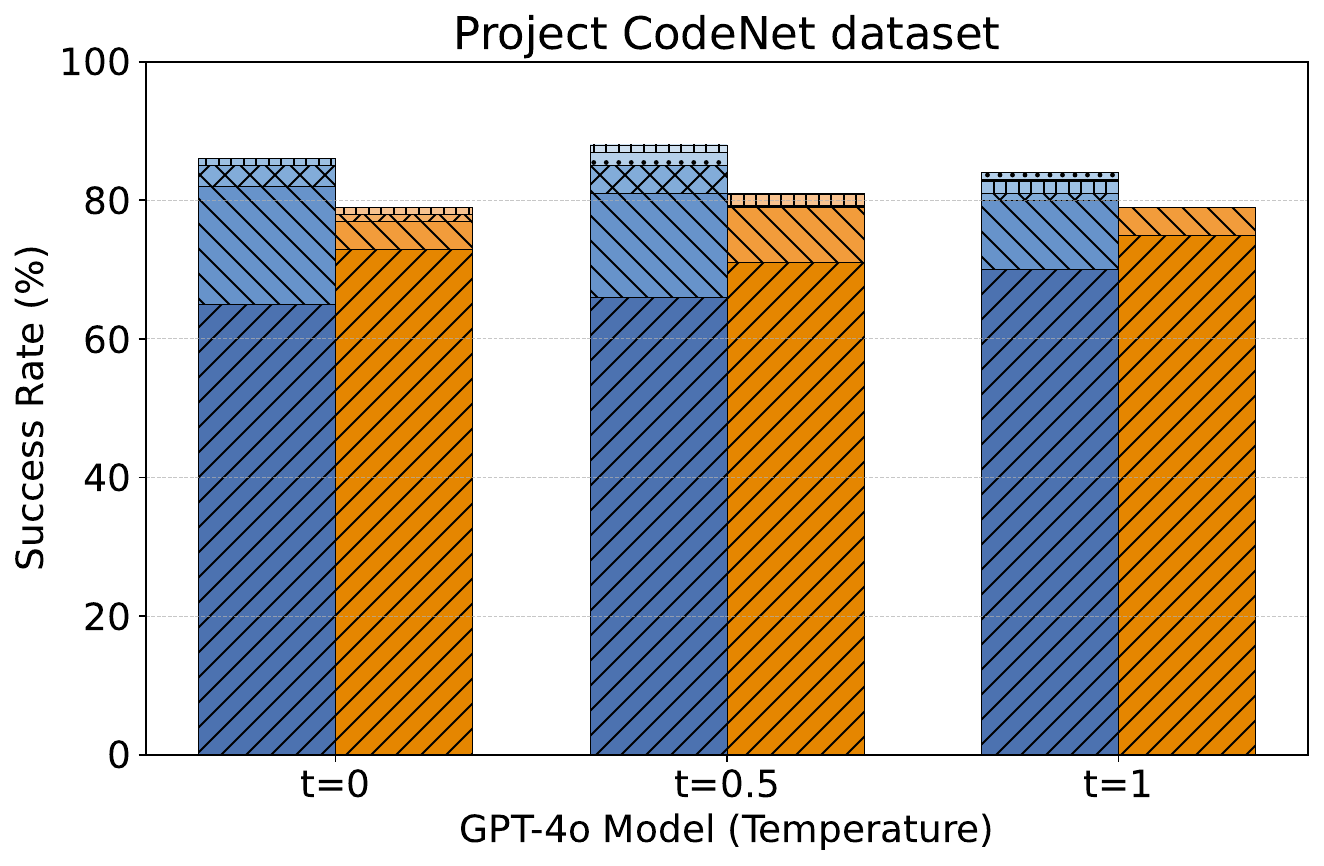}
		\caption{Success Rate on Project CodeNet}
		\label{fig:temp_codenet}
	\end{subfigure}
	\caption{Success Rate of \sys{} with different temperature settings for GPT-4o on TransCoder-IR and Project CodeNet datasets.}
	\label{fig:temperature}
\end{figure*}

In \S~\ref{sec:results}, all the experiments are conducted with the temperature set to default values, as explained on Appendix~\ref{app:llm_config}.
To investigate how temperature affects the performance of \sys{}, we conduct additional experiments with different temperature settings (0.0, 0.5, 1.0) for GPT-4o on both
TransCoder-IR and Project CodeNet datasets, as shown in Figure~\ref{fig:temperature}.
Through some preliminary experiments and discussions on OpenAI's community forum \footnote{https://community.openai.com/t/cheat-sheet-mastering-temperature-and-top-p-in-chatgpt-api/172683},
we find that setting the temperature more than 1 will likely to generate more random and less relevant outputs, which is not suitable for our task.

\textbf{(1) TransCoder-IR} (Figure~\ref{fig:temp_transcoder}): Setting the decoder to a deterministic temperature of $t = 0$ resulted in 83 successful translations (83\%), while both $t = 0.5$ and $t = 1.0$ yielded 80 successes (80\%) each. This represents a slightly improvement with 3 additional correct predictions under the deterministic setting.

\textbf{(2) Project CodeNet} (Figure~\ref{fig:temp_codenet}): Temperature does not have a significant impact: the model produced 79, 81, and 79 successful outputs at $t = 0$, $t = 0.5$, and $t = 1.0$ respectively (79--81\%), which does not indicate any outstanding trend in performance across the temperature settings.

The results on both datasets suggests that lowering temperature to zero can offer a slight boost in reliability
some of the cases, but it does not significantly affect the overall performance of \sys{}.

%% file: contents/appendix/app_N_spec_rules.tex
\section{Spec-driven Harness Rules}
\label{app:spec_rules}

\begin{figure}[h]
  \centering
  \includegraphics[width=0.7\linewidth]{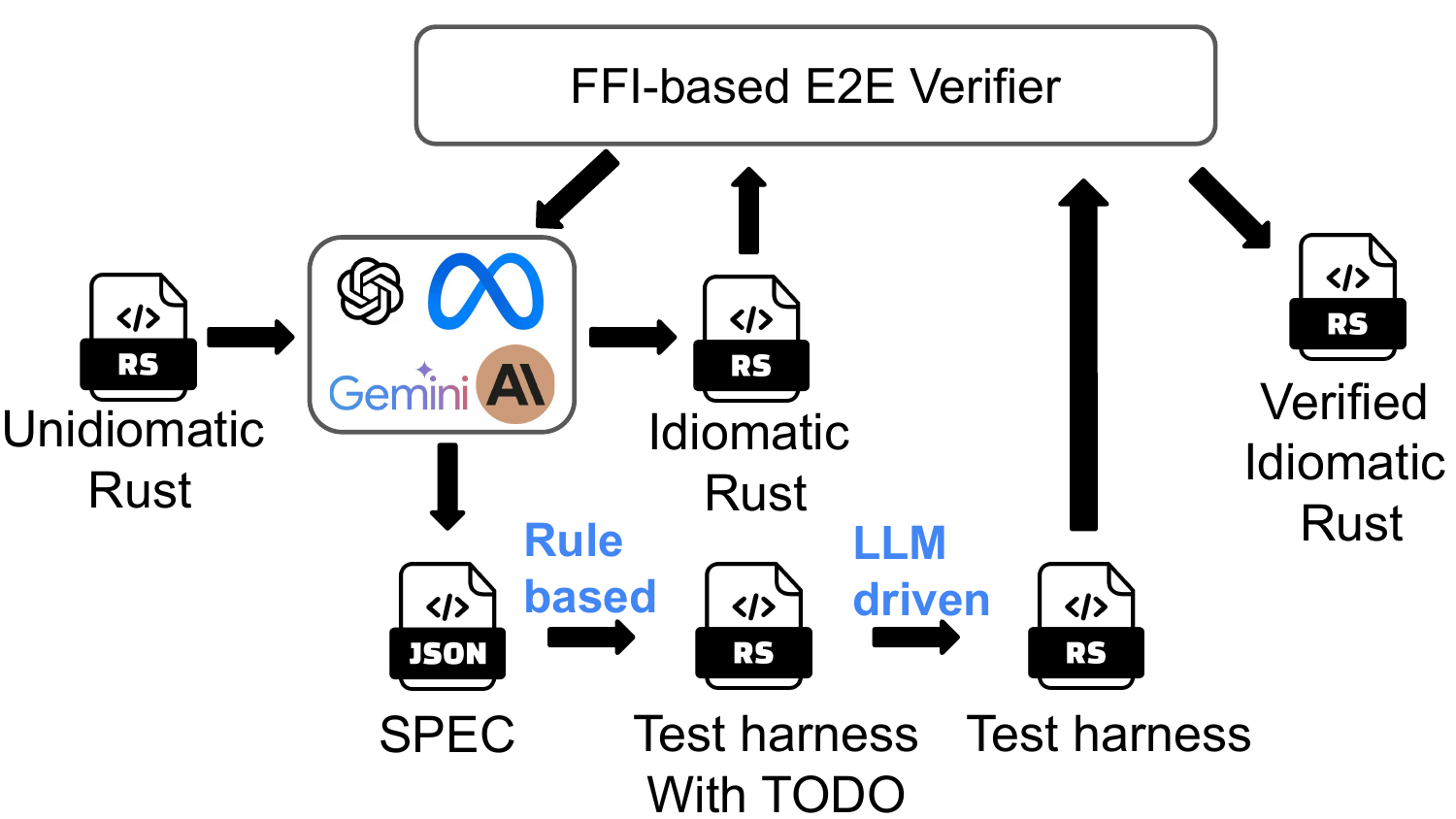}
  \caption{Spec-driven harness generation and verification loop. The idiomatic translator co-produces idiomatic Rust and a machine-readable SPEC. A rule-based generator synthesizes a C-compatible harness from the SPEC; unsupported mappings trigger a localized LLM fallback. Harness and idiomatic code are linked via FFI for end-to-end tests.}
  \label{fig:harness_gen}
\end{figure}

Figure~\ref{fig:harness_gen} illustrates the co-production timing and dataflow among artifacts (idiomatic code, SPEC, harness) and the verifier. Table~\ref{tab:spec_rules} summarizes the SPEC patterns our rule-based generator currently supports.

\input{figures/spec-table}

\paragraph{Harness construction details.}
The generator consumes a per-item SPEC (JSON) produced alongside idiomatic code and synthesizes: (i) a C-compatible shim that matches the original ABI, and (ii) idiomatic adapters that convert to/from Rust types. Pointer shapes (scalar, cstring, slice, ref) determine how memory is borrowed or owned; length sources come from sibling fields or constants; nullability and ownership hints select \texttt{Option<\,>} or strict checks. Return values are mapped back to U form, writing lengths when needed. This bridging resolves the ABI mismatch introduced by idiomatic function signatures.

\paragraph{Struct mappings and self-check.}
For structs, the SPEC defines bidirectional converters between unidiomatic and idiomatic layouts. We validate adapter consistency with a minimal roundtrip: Unidiomatic\(\rightarrow\)Idiomatic(1)\(\rightarrow\)Unidiomatic\(\rightarrow\)Idiomatic(2). The self-check compares Idiomatic(1) and Idiomatic(2) field-by-field according to \texttt{compare} hints: \texttt{by\_value} requires exact equality on scalar fields; \texttt{by\_slice} compares slice contents using the SPEC-recorded length source; \texttt{skip} omits fields that are aliasing views or externally owned to avoid false positives. Seed unidiomatic values are synthesized by an LLM guided by the SPEC so that nullability, ownership, and length sources are populated consistently.

\paragraph{Fallback and verification loop.}
When a SPEC uses patterns not yet implemented (e.g., pointer kinds outside \texttt{cstring}/\texttt{slice}/\texttt{ref}; non-trivial \texttt{len\_from} expressions; string args whose \texttt{spec.kind} $\neq$ \texttt{cstring}), the generator emits a localized TODO that is completed by an LLM using the same SPEC as guidance; the resulting harness is then validated as usual. End-to-end tests run against the linked harness and idiomatic crate; passing tests provide confidence under their coverage, while failures trigger the paper's feedback procedure for regeneration and refinement.

\subsection*{SPEC rule reference}
This section explains the rule families the SPEC uses to describe how unidiomatic, C-facing values become idiomatic Rust and back. The schema has two top-level forms: a struct description and a function description. Both are expressed as small collections of field mappings from the unidiomatic side to idiomatic paths; a function return is just another mapping whose idiomatic path is the special name \texttt{ret}. This uniform treatment keeps the generator simple and makes the SPEC readable by humans and machines alike.

Pointer handling is captured by a compact notion of shape. A field is either a scalar or one of three pointer shapes: a byte string that follows C conventions, a slice that pairs a pointer with a length, or a single-object reference. Slices record where their length comes from (either a sibling field or a constant). Each pointer also carries a null policy that distinguishes admissible NULL from forbidden NULL, which in turn selects idiomatic options versus strict checks in the generated adapters.

Two lightweight hints influence how the harness allocates and how the roundtrip self-check behaves. An ownership hint (owning vs transient) signals whether the idiomatic side should materialize owned data or borrow it for the duration of the call. A comparison hint (by value, by slice, or skip) declares how roundtrip checks should assert equality, so that aliasing views or externally owned buffers can be skipped without producing spurious failures.

Finally, the schema enforces well-formedness and defines a safe escape hatch. Invalid combinations are rejected early by validation. Patterns that are valid but not yet implemented by the generator, such as complex dotted paths or unusual pointer views, are localized and handed to the LLM fallback described earlier; the SPEC itself remains the single source of truth for the intended mapping.

%% file: figures/spec-table.tex
\begin{table*}[t]
\centering
\scriptsize
\setlength{\tabcolsep}{3pt}
\begin{tabularx}{\textwidth}{@{} >{\raggedright\arraybackslash}p{.14\textwidth}
                                    >{\raggedright\arraybackslash}p{.22\textwidth}
                                    >{\raggedright\arraybackslash}p{.22\textwidth}
                                    >{\raggedright\arraybackslash}p{.42\textwidth} @{}}
\toprule
\textbf{Category} & \textbf{SPEC keys} & \textbf{U (C/ABI) $\to$ I (Rust)} & \textbf{Notes / Status} \\
\midrule
Scalars & \texttt{shape: "scalar"} & scalar \(\rightarrow\) scalar & Common libc types are cast with \texttt{as} when needed; default compare is by value in roundtrip selftest. \\
\addlinespace[2pt]
C string & \texttt{ptr.kind: "cstring"}, \texttt{ptr.null} & \texttt{*const/*mut c\_char} \(\rightarrow\) \texttt{String} / \texttt{\&str} / \texttt{Option<String>} & NULL handling via \texttt{ptr.null} or \texttt{Option<\,>}; uses \texttt{CStr}/\texttt{CString} with lossless fallback. Return strings are converted back to \texttt{*mut c\_char}. \\
\addlinespace[2pt]
Slices & \texttt{ptr.kind: "slice"}, \texttt{len\_from}|\texttt{len\_const} & \texttt{*const/*mut T} + length \(\rightarrow\) \texttt{Vec<T>}, \texttt{\&[T]}, or \texttt{Option<...>} & Requires a length source; empty or NULL produces \texttt{None} or empty according to spec; writes back length on I\(\rightarrow\)U when a paired length field exists. \\
\addlinespace[2pt]
Single-element ref & \texttt{ptr.kind: "ref"} & \texttt{*const/*mut T} \(\rightarrow\) \texttt{Box<T>} / \texttt{Option<Box<T>>} & For struct \texttt{T}, generator calls auto struct converters \texttt{C\,T\_to\_T\_mut}/\texttt{T\_to\_C\,T\_mut}. \\
\addlinespace[2pt]
Derived length path & idiomatic path ending with \texttt{.len} & \texttt{len} field \(\leftrightarrow\) \texttt{vec.len} & Recognizes idiomatic \texttt{data.len} and reuses the same U-side length field on roundtrip. \\
\addlinespace[2pt]
Nullability & \texttt{ptr.null: nullable|forbidden} & C pointers \(\rightarrow\) field with/without \texttt{Option} & \texttt{nullable} maps to \texttt{Option<\,>} or tolerant empty handling. \\
\addlinespace[2pt]
\&mut struct params & ownership: transient & \texttt{*mut CStruct} \(\rightarrow\) \texttt{\&mut Struct} or \texttt{Option<\&mut Struct>} & Copies back mutated values after the call using generated struct converters. \\
\addlinespace[2pt]
Return mapping & Field with \texttt{i\_field.name = "ret"} & idiomatic return \(\rightarrow\) U output(s) & Scalars: direct or via \texttt{*mut T}. Strings: to \texttt{*mut c\_char}. Slices: pointer + length writeback. Structs: via struct converters. \\
\addlinespace[2pt]
Comparison hints & \texttt{compare: by\_value|by\_slice|skip} & selftest behavior & Optional per-field checks after U\(\rightarrow\)I1\(\rightarrow\)U\(\rightarrow\)I2 roundtrip, and compare with I1 and I2 \\
\addlinespace[2pt]
Unsupported paths & All SPEC key pairs other than supported paths & fallback & Generator emits localized TODOs for LLM completion; schema validation rejects malformed SPECs. \\
\bottomrule
\end{tabularx}
\caption{SPEC-driven harness coverage. U denotes the unidiomatic C-facing representation; I denotes the idiomatic Rust side.}
\label{tab:spec_rules}
\end{table*}

%% file: contents/appendix/app_O_crust_bench_failures.tex
\section{Real-world Codebase Evaluation Details}
\subsection{CRust-Bench Per-sample Outcomes}
\label{app:crust_failures}

\noindent
Table~\ref{tab:crust_failures} lists, for each of the 50 samples, the function-level translation status and a concise failure analysis. Status is reported as per-sample function-level percentages in separate columns for the unidiomatic (Unid.) and idiomatic (Id.) stages.

\subsection{\texttt{libogg} Outcomes}
\label{app:libogg-outcome}
\textit{(1) Using GPT-4o.} 36 functions cannot be translated idiomatically. nine of the translation failures are caused by translated functions not passing the test cases of \texttt{libogg}. Six failures are due to compile errors in the translations,
five of which result from the LLM violating Rust's safety rules on  lifetime, borrow, and mutability.
For example, the translation of function \texttt{\_os\_lacing\_expand} fails because the translation sets the value of a function parameter to a reference to the function's local  variable \texttt{vec}, leading to an error  ``\`{}vec\`{} does not live long enough."
Two failures are due to \sys{} being unable to generate compilable test harnesses.
If a function calls another function that \sys{} cannot translate, then the caller function cannot be translated either.  This is the reason why the remaining 13 translations fail.

\noindent\textit{(2) Using GPT-5.} 17 functions cannot be translated idiomatically. Among them, three are because the generated functions cannot pass the test cases  and three are due to failure to generate compilable test harnesses. Only one is caused by a compile error in the translated function, which shows the progress of GPT-5 in understanding Rust grammar and fixing compile errors. The remaining failures result from the  callee functions of those functions being untranslatable.

\input{figures/crust_bench_failure_table}

%% file: figures/crust_bench_failure_table.tex
\onecolumn
{\scriptsize
\begin{longtable}{ @{} p{.16\textwidth}@{} p{.06\textwidth}@{} p{.06\textwidth}@{} p{.45\textwidth} p{.25\textwidth} @{} }
\toprule
Sample & Unid. (\%) & Id. (\%) & Failure reason & Category \\
\midrule
\endfirsthead

\toprule
Sample & Unid. (\%) & Id. (\%) & Failure reason & Category \\
\midrule
\endhead
\midrule
\endfoot

\endlastfoot

\texttt{2DPartInt}& 100.0\% & 100.0\% & -- & -- \\
\texttt{42-Kocaeli-Printf}& 75.0\% & -- & C variadics require unstable \texttt{c\_variadic}; unresolved \texttt{va\_list} import blocks build. & Unidiomatic compile (C varargs/unstable feature) \\
\texttt{CircularBuffer}& 100.0\% & 54.6\% & CamelCase-to-snake\_case renaming breaks signature lookup; later run panics under no-unwind context. & Idiomatic compile (symbol/name mapping) \\
\texttt{FastHamming}& 100.0\% & 60.0\% & Output buffer sized to input length in harness; bounds-check panic at runtime. & Harness runtime (buffer/length) \\
\texttt{Holdem-Odds}& 100.0\% & 6.9\% & Off-by-one rank yields out-of-bounds bucket index; SIGSEGV under tests. & Runtime fault (boundary/indexing) \\
\texttt{Linear-Algebra-C}& 100.0\% & 44.8\% & Pointer vs reference semantics mismatch (nullable C pointers vs Rust references); harness compile errors. & Harness compile (pointer/ref semantics) \\
\texttt{NandC}& 100.0\% & 100.0\% & -- & -- \\
\texttt{Phills\_DHT}& 75.0\% & -- & Shadowed global \texttt{hash\_table} keeps \texttt{dht\_is\_initialised()} false; assertion in tests. & Runtime fault (global state divergence) \\
\texttt{Simple-Sparsehash}& 100.0\% & 40.0\% & CamelCase-to-snake\_case renaming causes signature/type mismatches; harness does not compile. & Idiomatic compile (symbol/name mapping) \\
\texttt{SimpleXML}& 83.3\% & -- & Missing \texttt{ParseState} and CamelCase-to-snake\_case renaming breaks signatures; unidiomatic stalls. & Idiomatic compile (symbol/name mapping) \\
\texttt{aes128-SIMD}& 85.7\% & -- & Array-shape mismatch (expects 4x4 refs; passes row pointer); plus intrinsics/typedef noise. & Unidiomatic compile (array shape; intrinsics/types) \\
\texttt{amp}& 80.0\% & -- & Returned C string from \texttt{amp\_decode\_arg} is not NULL-terminated; \texttt{strcmp} reads past allocation and trips invalid read under tests. & Runtime fault (C string NULL termination) \\
\texttt{approxidate}& 85.7\% & -- & \texttt{match\_alpha} references anonymous enum \texttt{C2RustUnnamed} that is never defined, causing cascaded missing-type errors across retries. & Unidiomatic compile (types/aliases) \\
\texttt{avalanche}& 100.0\% & 75.0\% & Capturing closure passed where \texttt{fn} pointer required; FILE*/Rust \texttt{File} bridging mis-modeled; compile fails. & Harness runtime (I/O/resource model mismatch) \\
\texttt{bhshell}& 88.2\% & -- & Many parser errors (enum lacks \texttt{PartialEq}, missing consts, u64 to usize drift, duplicates). & Unidiomatic compile (types/aliases) \\
\texttt{bitset}& 100.0\% & 50.0\% & Treats bit count as byte count in converter; overreads and panics under tests. & Harness runtime (buffer/length) \\
\texttt{bostree}& 52.4\% & -- & Function-pointer typedefs and pointer-shape drift break callback bridging. & Unidiomatic compile (function-pointer types/deps) \\
\texttt{btree-map}& 100.0\% & 26.3\% & Trace/instrumentation proc macro requires \texttt{Debug} on opaque C type \texttt{node}; harness compilation fails for \texttt{get\_node\_count}. & Harness compile (instrumentation bound) \\
\texttt{c-aces}& 100.0\% & 3.9\% & Struct converter mismatch (\texttt{Vec<CMatrix2D>} vs \texttt{Vec<Matrix2D>}) in generated harness; compile fails after retries. & Harness compile (struct converter/shape) \\
\texttt{c-string}& 100.0\% & 29.4\% & Size vs capacity mismatch in \texttt{StringT} constructor; empty buffer returned, C asserts. & Runtime fault (size/capacity mismatch) \\
\texttt{carrays}& 100.0\% & 68.5\% & Trace macro imposes \texttt{Debug} on generic \texttt{T} and callback; harness fails to compile (e.g., \texttt{gca\_lsearch}). & Harness compile (instrumentation bound) \\
\texttt{cfsm}& 50.0\% & -- & Missing typedefs for C function-pointer callbacks; harness lacks nullable extern signatures, compile fails. & Unidiomatic compile (function-pointer types/deps) \\
\texttt{chtrie}& 100.0\% & 0.0\% & Pointer-of-pointers vs \texttt{Vec} adapter mismatch for struct \texttt{chtrie} & Harness compile (struct converter/shape) \\
\texttt{cissy}& 100.0\% & 19.1\% & Anonymous C types that c2rust renamed cannot be fetched correctly as a dependency & Unidiomatic compile (types/aliases) \\
\texttt{clog}& 31.6\% & -- & Variadic logging APIs and duplicate globals; unresolved \texttt{vfprintf}/\texttt{c\_variadic}; compile fails. & Unidiomatic compile (C varargs/unstable feature) \\
\texttt{cset}& 100.0\% & 25.0\% & Translator renames \texttt{XXH\_readLE64} to \texttt{xxh\_read\_le64}; SPEC/harness require exact C name; exhausts six attempts. & Idiomatic compile (symbol/name mapping) \\
\texttt{csyncmers}& 66.7\% & -- & Unsigned underflow in \texttt{compute\_closed\_syncmers} (\texttt{i - S + 1} without guard) triggers overflow panic; prior \texttt{\_\_uint128\_t} typedef issues. & Runtime fault (arithmetic underflow) \\
\texttt{dict}& 17.7\% & -- & Fn-pointer fields modeled non-optional (need \texttt{Option<extern "C" fn>}); plus \texttt{va\_list} requires nightly \texttt{c\_variadic}; compile fails. & Unidiomatic compile (function-pointer types/deps) \\
\texttt{emlang}& 16.3\% & -- & Anonymous-union alias (\texttt{C2RustUnnamed}) misuse; duplicate \texttt{program\_new}; assertion bridging (\texttt{\_\_assert\_fail}) mis-modeled. & Unidiomatic compile (types/aliases) \\
\texttt{expr}& 33.3\% & -- & Missing \texttt{C2RustUnnamed} alias; C varargs in \texttt{trace\_eval}; \texttt{strncmp} len type mismatch. & Unidiomatic compile (types/aliases) \\
\texttt{file2str}& 100.0\% & 100.0\% & -- & -- \\
\texttt{fs\_c}& 100.0\% & 60.0\% & Idiomatic I/O wrappers mismatch C expectations (closed fd/OwnedFd abort; Err(NotFound) leads to C-side segfault). & Harness runtime (I/O/resource model mismatch) \\
\texttt{geofence}& 100.0\% & 100.0\% & -- & -- \\
\texttt{gfc}& 100.0\% & 54.6\% & Converter overread + ownership misuse; later compile errors. & Harness runtime (converter/ownership) \\
\texttt{gorilla-paper-encode}& 100.0\% & 9.1\% & Missing adapters + lifetimes (\texttt{Cbitwriter\_s}/\texttt{Cbitreader\_s} vs \texttt{BitWriter}/\texttt{BitReader<'a>}). & Harness compile (lifetimes/struct adapters) \\
\texttt{hydra}& 100.0\% & 50.0\% & Borrow overlap in list update; name mapping for \texttt{FindCommand}. & Idiomatic compile (borrow/lifetime; symbol mapping) \\
\texttt{inversion\_list}& 17.0\% & -- & C allows NULL comparator/function pointers; wrapper unwraps and panics. & Runtime fault (function-pointer nullability) \\
\texttt{jccc}& 88.7\% & -- & Missing \texttt{C2RustUnnamed} alias and duplicate \texttt{Expression}/\texttt{Lexer} types; compile fails. & Unidiomatic compile (types/aliases) \\
\texttt{leftpad}& 100.0\% & 100.0\% & -- & -- \\
\texttt{lib2bit}& 100.0\% & 13.6\% & Non-clonable \texttt{std::fs::File} in harness (C \texttt{FILE*} vs Rust \texttt{File} I/O handle mismatch) & Harness runtime (I/O/resource model mismatch) \\
\texttt{libbase122}& 100.0\% & 37.5\% & Reader cursor/buffer not preserved across calls; writer shape mismatch; tests fail. & Harness runtime (converter/ownership) \\
\texttt{libbeaufort}& 100.0\% & 66.7\% & Returns reference to temporary tableau; matrix parameter shape drift (\texttt{char**} vs \texttt{Vec<Option<String>>}); compile fails. & Idiomatic compile (borrow/lifetime) \\
\texttt{libwecan}& 100.0\% & 100.0\% & -- & -- \\
\texttt{morton}& 100.0\% & 100.0\% & -- & -- \\
\texttt{murmurhash\_c}& 100.0\% & 100.0\% & -- & -- \\
\texttt{razz\_simulation}& 33.3\% & -- & Type-name drift; node shape; ptr/ref API mismatch. & Harness compile (type/name drift; API mismatch) \\
\texttt{rhbloom}& 100.0\% & 33.3\% & Pointer/ref misuse; bit-length as bytes; overreads/panics. & Harness runtime (pointer/ref; length units) \\
\texttt{totp}& 77.8\% & -- & Anonymous C types that c2rust renamed cannot be fetched correctly as a dependency; plus duplicate helpers (\texttt{pack32}/\texttt{unpack64}/\texttt{hmac\_sha1}); compile fails. & Unidiomatic compile (types/aliases) \\
\texttt{utf8}& 100.0\% & 30.8\% & NULL deref + unchecked indices; SIGSEGV in tests. & Runtime fault (NULL deref/out-of-bounds) \\
\texttt{vec}& 100.0\% & 0.0\% & Idiomatic rewrite uses a bounds-checked copy; out-of-range panic under tests. & Runtime fault (boundary/indexing) \\
\bottomrule
\caption{CRust-Bench per-sample outcomes (function-level). Translation Status columns report per-sample function-level success rates for unidiomatic (Unid.) and idiomatic (Id.) stages.}\label{tab:crust_failures}\\
\end{longtable}
}

%% file: contents/appendix/app_M_prompts.tex
\section{Examples of Prompts Used in \sys{}}
\label{app:prompts}

The following prompts are used to guide the LLM in C-to-Rust translation and
verification tasks. The prompts may slightly vary to accommodate different
translation task, as \sys{} leverages static
analysis to fetch the necessary information for the LLM.

\subsection{Unidiomatic Translation}

Figure~\ref{fig:unidiomatic-translation} shows the prompt for translating
unidiomatic C code to Rust.

\captionsetup{type=figure}
\lstset{basicstyle=\ttfamily\scriptsize,breaklines=true}
	\begin{lstlisting}[language={}]
Translate the following C function to Rust. Try to keep the **equivalence** as much as possible.
`libc` will be included as the **only** dependency you can use. To keep the equivalence, you can use `unsafe` if you want.
The function is:
```c
{C_FUNCTION}
```

// Specific for main function
The function is the `main` function, which is the entry point of the program. The function signature should be: `pub fn main() -> ()`.
For `return 0;`, you can directly `return;` in Rust or ignore it if it's the last statement.
For other return values, you can use `std::process::exit()` to return the value.
For `argc` and `argv`, you can use `std::env::args()` to get the arguments.

The function uses some of the following stdio file descriptors: stdin. Which will be included as
```rust
extern "C" {
    static mut stdin: *mut libc::FILE;
}

```
You should **NOT** include them in your translation, as the system will automatically include them.

The function uses the following functions, which are already translated as (you should **NOT** include them in your translation, as the system will automatically include them):
```rust
{DEPENDENCIES}
```

Output the translated function into this format (wrap with the following tags):
----FUNCTION----
```rust
// Your translated function here
```
----END FUNCTION----
\end{lstlisting}
\captionof{figure}{Unidiomatic Translation Prompt}\label{fig:unidiomatic-translation}

\subsection{Unidiomatic Translation with Feedback}

Figure~\ref{fig:unidiomatic-translation-feedback} shows the prompt for
translating unidiomatic C code to Rust with feedback from the previous incorrect
translation and error message.

\captionsetup{type=figure}
\lstset{basicstyle=\ttfamily\scriptsize,breaklines=true}
	\begin{lstlisting}[language={}]
Translate the following C function to Rust. Try to keep the **equivalence** as much as possible.
`libc` will be included as the **only** dependency you can use. To keep the equivalence, you can use `unsafe` if you want.
The function is:
```c
{C_FUNCTION}

```

// Specific for main function
The function is the `main` function, which is the entry point of the program. The function signature should be: `pub fn main() -> ()`.
For `return 0;`, you can directly `return;` in Rust or ignore it if it's the last statement.
For other return values, you can use `std::process::exit()` to return the value.
For `argc` and `argv`, you can use `std::env::args()` to get the arguments.

The function uses some of the following stdio file descriptors: stdin. Which will be included as
```rust
extern "C" {
    static mut stdin: *mut libc::FILE;
}

```
You should **NOT** include them in your translation, as the system will automatically include them.

The function uses the following functions, which are already translated as (you should **NOT** include them in your translation, as the system will automatically include them):
```rust
fn atoi (str : * const c_char) -> c_int;
```

Output the translated function into this format (wrap with the following tags):
----FUNCTION----
```rust
// Your translated function here
```
----END FUNCTION----

Lastly, the function is translated as:
```rust
{COUNTER_EXAMPLE}
```
It failed to compile with the following error message:
```
{ERROR_MESSAGE}
```
Analyzing the error messages, think about the possible reasons, and try to avoid this error.
\end{lstlisting}
\captionof{figure}{Unidiomatic Translation with Feedback Prompt}\label{fig:unidiomatic-translation-feedback}

\subsection{Idiomatic Translation}

Figure~\ref{fig:idiomatic-translation} shows the prompt for translating
unidiomatic Rust code to idiomatic Rust. Crown is used to hint the LLM about the
ownership, mutability, and fatness of pointers.

\captionsetup{type=figure}
\begin{lstlisting}[language={},basicstyle=\ttfamily\scriptsize,breaklines=true]
Translate the following unidiomatic Rust function into idiomatic Rust. Try to remove all the `unsafe` blocks and only use the safe Rust code or use the `unsafe` blocks only when necessary.
Before translating, analyze the unsafe blocks one by one and how to convert them into safe Rust code.
**libc may not be provided in the idiomatic code, so try to avoid using libc functions and types, and avoid using `std::ffi` module.**
```rust
{RUST_FUNCTION}
```

"Crown" is a pointer analysis tool that can help to identify the ownership, mutability and fatness of pointers. Following are the possible annotations for pointers:
```
fatness:
    - `Ptr`: Single pointer
    - `Arr`: Pointer is an array
mutability:
    - `Mut`: Mutable pointer
    - `Imm`: Immutable pointer
ownership:
    - `Owning`: Owns the pointer
    - `Transient`: Not owns the pointer
````

The following is the output of Crown for this function:
```
{CROWN_RESULT}
```
Analyze the Crown output firstly, then translate the pointers in function arguments and return values with the help of the Crown output.
Try to avoid using pointers in the function arguments and return values if possible.

Output the translated function into this format (wrap with the following tags):
----FUNCTION----
```rust
// Your translated function here
```
----END FUNCTION----

Also output a minimal JSON spec that maps the unidiomatic Rust layout to the idiomatic Rust for the function arguments and return value.
Full JSON Schema for the SPEC (do not output the schema; output only an instance that conforms to it):
```json
{_schema_text}
```
----SPEC----
```json
{{
  "function_name": "{function.name}",
  "fields": [
    {{
      "u_field": {{
        "name": "...",
        "type": "...",
        "shape": "scalar" | {{"ptr": {{"kind": "slice|cstring|ref", "len_from": "?", "len_const": 1}}}}
      }},
      "i_field": {{
        "name": "...",
        "type": "..."
      }}
    }}
  ]
}}
```
----END SPEC----
Few-shot examples (each with unidiomatic Rust signature, idiomatic Rust signature, and the SPEC):
Example F1 (slice arg):
Unidiomatic Rust:
```rust
pub unsafe extern "C" fn sum(xs: *const i32, n: usize) -> i32;
```
Idiomatic Rust:
```rust
pub fn sum(xs: &[i32]) -> i32;
```
----SPEC----
```json
{{
  "function_name": "sum",
  "fields": [
    {{ "u_field": {{"name": "xs", "type": "*const i32", "shape": {{ "ptr": {{ "kind": "slice", "len_from": "n" }} }} }},
       "i_field": {{"name": "xs", "type": "&[i32]" }} }},
    {{ "u_field": {{"name": "n",  "type": "usize",      "shape": "scalar" }},
       "i_field": {{"name": "xs.len", "type": "usize" }} }}
  ]
}}
```
----END SPEC----

Example F2 (ref out):
Unidiomatic Rust:
```rust
pub unsafe extern "C" fn get_value(out_value: *mut i32);
```
Idiomatic Rust:
```rust
pub fn get_value() -> i32;
```
----SPEC----
```json
{{
  "function_name": "get_value",
  "fields": [
    {{ "u_field": {{"name": "out_value", "type": "*mut i32", "shape": {{ "ptr": {{ "kind": "ref" }} }} }},
       "i_field": {{"name": "ret", "type": "i32" }} }}
  ]
}}
```
----END SPEC----

Example F3 (nullable cstring maps to Option):
Unidiomatic Rust:
```rust
pub unsafe extern "C" fn set_name(name: *const libc::c_char);
```
Idiomatic Rust:
```rust
pub fn set_name(name: Option<&str>);
```
----SPEC----
```json
{{
  "function_name": "set_name",
  "fields": [
    {{ "u_field": {{"name": "name", "type": "*const c_char", "shape": {{ "ptr": {{ "kind": "cstring", "null": "nullable" }} }} }},
       "i_field": {{"name": "name", "type": "Option<&str>" }} }}
  ]
}}
```
----END SPEC----
\end{lstlisting}
\captionof{figure}{Idiomatic Translation Prompt}\label{fig:idiomatic-translation}

\subsection{Idiomatic Verification}

Idiomatic verification is the process of verifying the correctness of the
translated idiomatic Rust code by generating a test harness. The prompt for
idiomatic verification is shown in Figure~\ref{fig:idiomatic-verification}.

\captionsetup{type=figure}
\lstset{basicstyle=\ttfamily\scriptsize,breaklines=true}
\begin{lstlisting}[language={}]

We have an initial spec-driven harness with TODOs. Finish all TODOs and ensure it compiles.
Idiomatic signature:
```rust
pub fn compute_idiomatic(
    x: i32,
    name: &str,
    data: &[u8],
    meta: HashMap<String, String>,
) -> i32;;
```
Unidiomatic signature:
```rust
pub unsafe extern "C" fn compute(x: i32, name: *const libc::c_char, data: *const u8, len: usize, meta: *const libc::c_char) -> i32;;
```

Current harness:
```rust
pub unsafe extern "C" fn compute(x: i32, name: *const libc::c_char, data: *const u8, len: usize, meta: *const libc::c_char) -> i32
{
    // Arg 'name': borrowed C string at name
    let name_str = if !name.is_null() {
        unsafe { std::ffi::CStr::from_ptr(name) }.to_string_lossy().into_owned()
    } else {
        String::new()
    };
    // Arg 'data': slice from data with len len as usize
    let data_len = len as usize;
    let data_len_non_null = if data.is_null() { 0 } else { data_len };
    let data: &[u8] = if data_len_non_null == 0 {
        &[]
    } else {
        unsafe { std::slice::from_raw_parts(data as *const u8, data_len_non_null) }
    };
    // TODO: param meta of type HashMap < String , String >: unsupported mapping
    let __ret = compute_idiomatic(x, &name_str, data, /* TODO param meta */);
    return __ret;
}
```
Output only the final function in this format:
----FUNCTION----
```rust
// Your translated function here
```
----END FUNCTION----
\end{lstlisting}
\captionof{figure}{Idiomatic Verification Prompt}\label{fig:idiomatic-verification}

\subsection{Failure Reason Analysis}
\label{app:failure-reason-analysis}

Figure~\ref{fig:failure-reason-analysis} shows the prompt for analyzing the
reasons for the failure of the translation.

\captionsetup{type=figure}
\lstset{basicstyle=\ttfamily\scriptsize,breaklines=true}
\begin{lstlisting}[language={}]
Given the following C code:
```c
{original_code}
```
The following code is generated by a tool that translates C code to Rust code. The tool has a bug that causes it to generate incorrect Rust code. The bug is related to the following error message:
```json
{json_data}
```
Please analyze the error message and provide a reason why the tool generated incorrect Rust code.

1. Append a new reason to the list of reasons.
2. Select a reason from the list of reasons that best describes the error message.

Please provide a reason why the tool generated incorrect Rust code **FUNDAMENTALLY**.

List of reasons:
{all_current_reasons}

Please provide the analysis output in the following format:
```json
{
    "action": "append", // or "select" to select a reason from the list of reasons
    "reason": "Format string differences between C and Rust", // the reason for the error message, if action is "append"
    "selection": 1 // the index of the reason from the list of reasons, if action is "select"
    // "reason" and "selection" are mutually exclusive, you should only provide one of them
}
```

Please **make sure** to provide a general reason that can be applied to multiple cases, not a specific reason that only applies to the current case.
Please provide a reason why the tool generated incorrect Rust code **FUNDAMENTALLY** (NOTE that the reason of first failure is always NOT the fundamental reason).
\end{lstlisting}
\captionof{figure}{Failure Reason Analysis Prompt}\label{fig:failure-reason-analysis}